\documentclass[a4paper,pre,reqno,superscriptaddress,twocolumn,showpacs,footinbib]{revtex4-1}

\usepackage{graphicx,color}
\usepackage{dcolumn}
\usepackage{epsfig}
\usepackage{hyperref}
\usepackage[centertags]{amsmath}
\usepackage{amsfonts,amsmath}
\usepackage{euscript}
\usepackage{amssymb}
\usepackage{amsthm}
\usepackage{newlfont}
\usepackage{mathrsfs}
\usepackage{subfigure}

\newcommand{\opunit}{\text{1}\kern-0.22em\text{l}}

\newcommand{\eg}{\textit{e.g.}}
\newcommand{\ie}{\textit{i.e.}}

\newcommand{\id}{\textrm{d}}

\newcommand{\beq}{\begin{equation}}
\newcommand{\eeq}{\end{equation}}

\def\bea{\begin{eqnarray}}
\def\eea{\end{eqnarray}}
\def\ba{\begin{array}}
\def\ea{\end{array}}
\def\n{\nonumber}

\def\la{\langle}
\def\ra{\rangle}

\def \al{\alpha}

\begin{document}

\title{On the universal Gaussian behavior of Driven Lattice Gases at short-times}

\author{Valerio Volpati}
\affiliation{IPhT -- Institut de Physique Th\'{e}orique, Universit\'{e} Paris Saclay, CEA, CNRS, F-91191 Gif-sur-Yvette}
\author{Urna Basu}
\affiliation{SISSA -- International School for Advanced Studies and INFN, via Bonomea 265, I-34136 Trieste}
\author{Sergio Caracciolo}
\affiliation{Universit\`a degli Studi di Milano -- Dip.~di Fisica and INFN, via Celoria 16, I-20133 Milano}
\author{Andrea Gambassi}
\affiliation{SISSA -- International School for Advanced Studies and INFN, via Bonomea 265, I-34136 Trieste}

\begin{abstract}
 The dynamic and static critical behaviors of driven and equilibrium lattice gas models are studied in two spatial dimensions. We show that in the short-time regime immediately following  a critical quench, the dynamics of the transverse order parameters,  auto-correlations, and Binder cumulant are consistent with the prediction of a Gaussian, $i.e.,$ non-interacting, effective theory, both for the equilibrium lattice gas and its nonequilibrium counterparts. 
Such a ``super-universal'' behavior is observed only at short times after a critical quench, while  the various models display their distinct behaviors in the stationary states, described by the corresponding, known universality classes.

\end{abstract}


\maketitle

\section{Introduction}

The search for universal behavior, which unites a class of systems in terms of some common collective properties, lies at the very heart of Statistical Physics.  Both in its static and dynamic manifestations, universality emerges  in large systems of interacting degrees of freedom close to a critical point, when they display a behavior which is actually independent of their microscopic features. This critical behavior is usually dictated by symmetry properties, or conservation laws; critical phenomena occurring in various systems having the same symmetries belong to the same universality class. Beyond its numerous and celebrated manifestations in equilibrium  \cite{Kadanoff}, universality plays an important role also in the dynamical relaxation of nonequilibrium systems \cite{TauberBook}, ranging from diffusive \cite{HalperinHohenberg1977} and reaction-diffusion \cite{DP} systems to surface growth \cite{KPZ}.

Remarkably, universality emerges not only in systems that are close to their stationary state, but also far from it, $i.e.,$ during the early stages  of the relaxation process, when the correlation length of the fluctuations of the relevant order parameter is still very small compared to the system size \cite{JSS, Zheng}. This fact often translates in the observation of novel critical exponents, but also in the possibility to measure the equilibrium and dynamical critical exponents which characterize the stationary state from the observation of this nonequilibrium relaxation, with a substantial reduction of the numerical costs \cite{AS2002,ASreview,Tauber}.

Classifying and characterizing nonequilibrium universality classes remain a challenge in Statistical Physics. Investigations of lattice models 
are very useful in this respect:  their simplicity makes them amenable to numerical, and sometimes analytical studies, yet they often show rich and novel physical phenomena. Lattice gases, which describe stochastic hopping of particles on a lattice, belong to one such class of models which has been extensively used to explore critical phenomena in and out of equilibrium \cite{Dickman,DDSbook}. These systems often show a continuous transition to an ordered state, where the particles cluster together, at a certain critical temperature. Such transitions are relevant in various physical situations including binary mixtures \cite{HalperinHohenberg1977}, driven diffusive systems \cite{DDSbook}, viscoelastic fluids \cite{visco}, vehicular traffic \cite{traffic} and active matter \cite{active2, Cates}.

The critical behavior characterizing the phase transitions in the various lattice gas models depends on the symmetries of their specific dynamics: equilibrium \cite{Mussardo}, driven  \cite{KLS, DDSbook} and randomly driven  \cite{rdlg-S} lattice gases therefore belong to different universality classes. All these models have one common feature though: the density of particles is locally conserved by the dynamics. Such a conservation law strongly constrains and slows the dynamics down, hence different dynamical behaviors are expected and observed compared to  non-conserved models \cite{TauberBook}.  While considerable amount of work has been devoted to study the critical behaviors of the latter, both in and out of equilibrium, much less attention has been given to the short-time dynamics of the former.

In this work we show that, remarkably, a sort of  ``super-universality'', which unites the different lattice gas models, emerges in the short-time regime after a critical quench,  irrespective of their specific critical behavior. In this regime, in fact,  the dynamical behavior of certain observables which can be considered the natural order parameters for these transitions are described by a non-interacting (Gaussian) effective theory.  
In particular, we will focus on the behavior of ``transverse'' observables in the driven lattice gas, the randomly driven lattice gas and the equilibrium lattice gas in two spatial dimensions.   
Despite the fact that features like the driving or the spatial anisotropy introduce a relevant perturbations in the lattice gases which change entirely the critical properties of the system, the short-time behavior of these natural observables is independent of these features.  The peculiar properties of specific universality classes are recovered, for all observables, only at longer times. 

The presentation is organized as follows: In Sec. \ref{sec:models} we recall the equilibrium and driven lattice gas models and define certain relevant observables. A brief discussion of the different effective field theories introduced in the past in order to study the critical behaviors of these models is presented in Sec. \ref{sec:ph-th}. Based on a Gaussian theory, the dynamical behavior of the transverse order parameters and auto-correlation of one of them are computed. This section elaborates and substantially extends the analysis of Ref.~\cite{prl-dlg}. 
In Sec. \ref{sec:short-t}  we compare the results obtained from Monte Carlo simulations in the short-time regime with the predictions of a Gaussian effective theory for both the driven and equilibrium lattice gases. The time evolution of the Binder cumulant 
starting from various initial states is also studied in the various models. Section \ref{sec:long-t} is devoted to the study of the stationary state behavior of the conserved lattice gases. 
We conclude with some general remarks in Sec. \ref{sec:concl}.

\section{The models}\label{sec:models}

We consider a periodic $d-$dimensional hyper-cubic lattice with size $V = L_\| \times L_\perp^{d-1}.$ 
The generic $i$-th site of the lattice can be either empty or occupied by a particle with a corresponding occupation number $n_i=0,1.$ The particles interact via a nearest-neighbour Ising Hamiltonian,
\begin{equation}
\mathcal{H}(C) = -4 \sum_{\langle i,j \rangle} n_i n_j,
\end{equation}
which depends on the configuration $C= \{ n_1, n_2, \dots n_V \}.$  We consider the case of a half-filled lattice, $i.e.,$ the total number $\sum_i n_i$ of particles is  fixed to be $ V/2.$ 

\begin{figure}[h]
 \centering
 \includegraphics[width=4.2 cm]{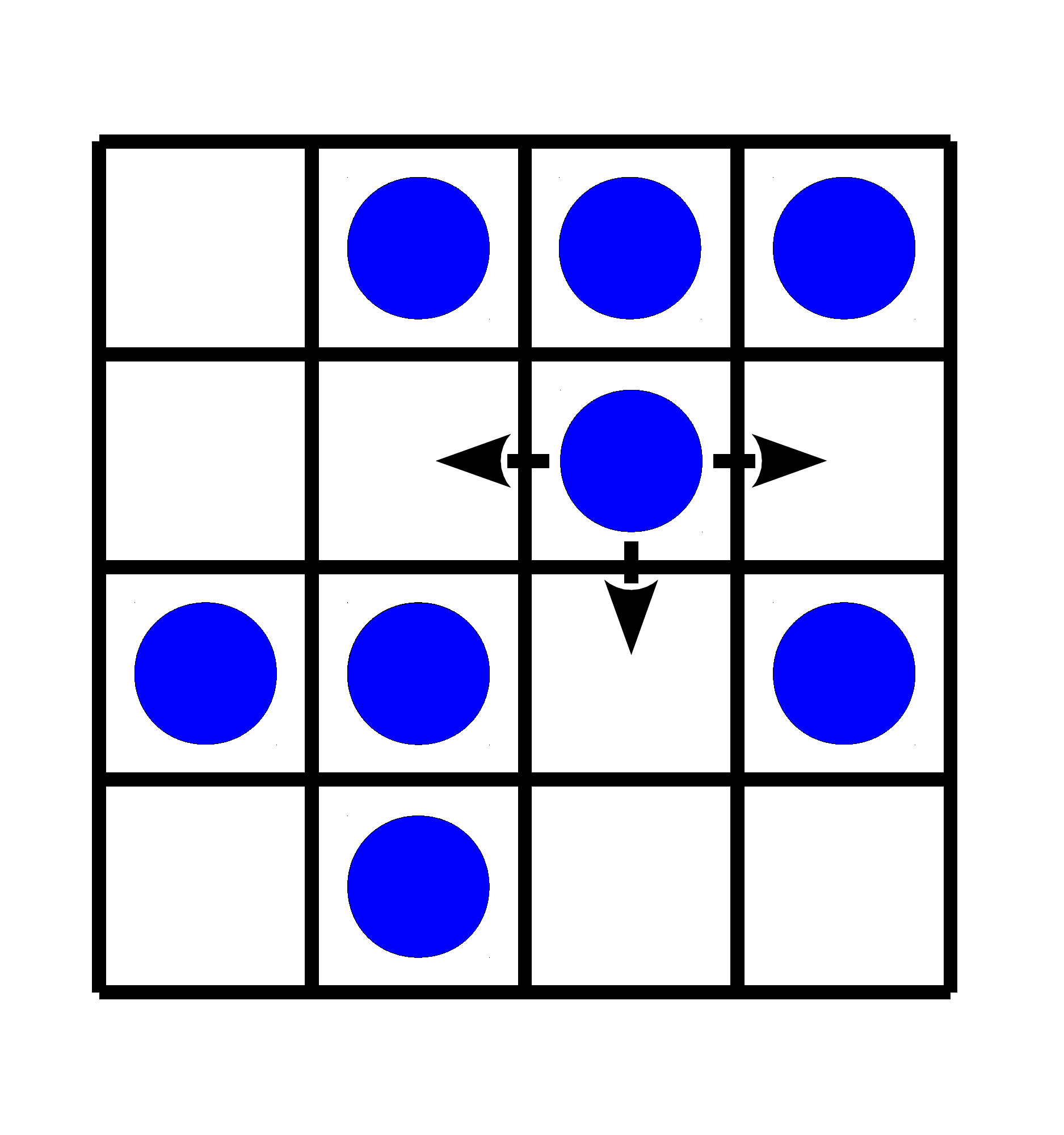} 
 \includegraphics[width=4.2 cm]{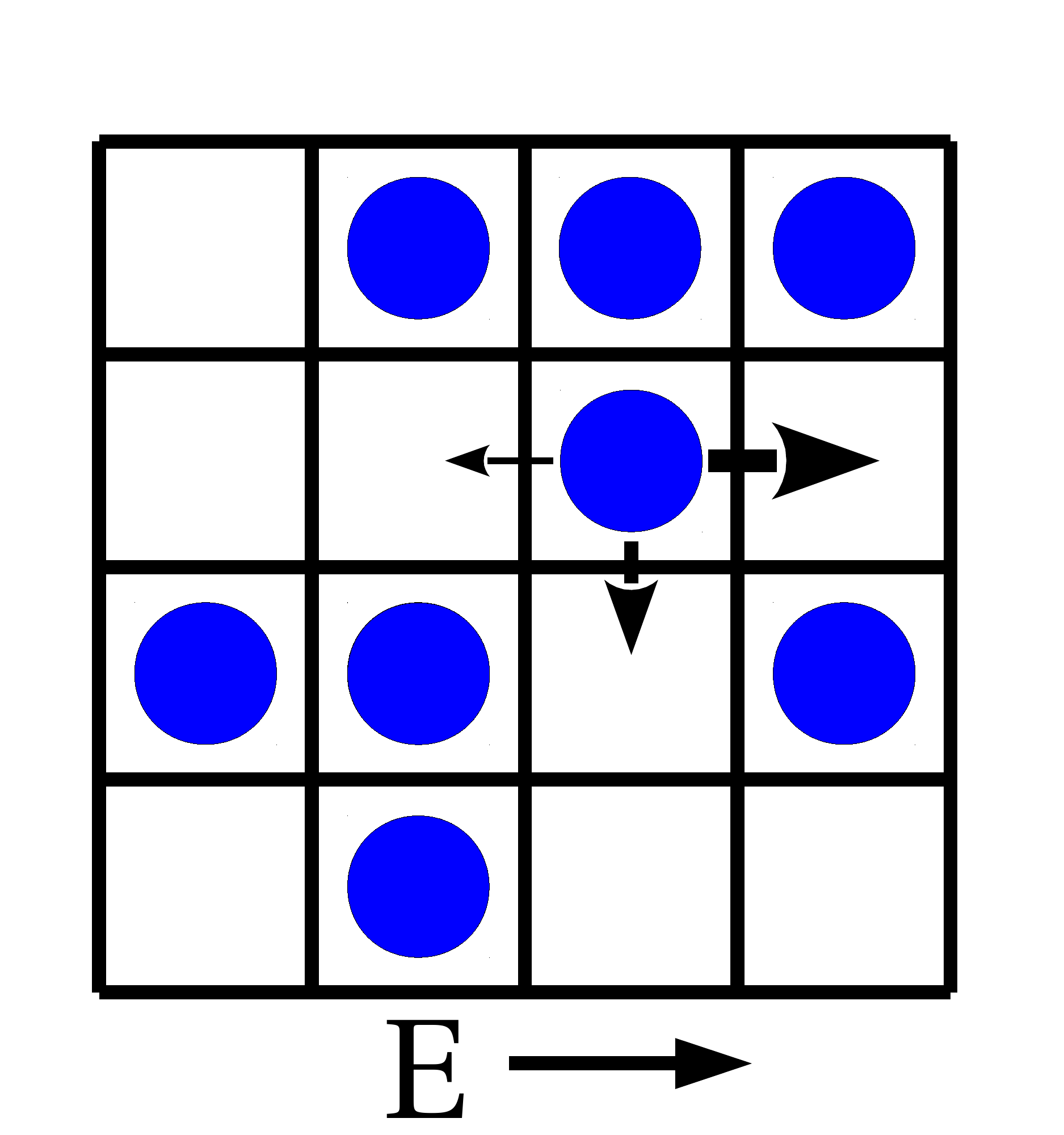} 
  \caption{Cartoon of the  LG (left) and DLG (right) dynamics on a two-dimensional lattice. In the LG, the  transition probability for the selected particle to jump in any of its empty  neighbouring sites depends only on the number of occupied neighbour sites and on the temperature. In the DLG, jumps are biased in the direction of the field $E$, such that for the selected particle there is a larger probability to jump along the field and a smaller probability to jump against it. }
 \label{fig:cartoon_dlg}
\end{figure}

\begin{figure*}[th]
 \centering
 \includegraphics[width=3.5 cm]{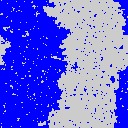} \hspace*{0.9 cm} 
 \includegraphics[width=3.5 cm]{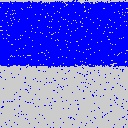} \hspace*{0.9 cm} 
 \includegraphics[width=3.5 cm]{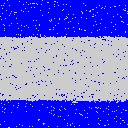}  
  \caption{Snapshots of the typical low-temperature (ordered) configurations of the LG (left), IDLG (centre) and RDLG (right) at half-filling and in the stationary state as obtained from the Monte Carlo simulation on square lattices of size 128 $\times$ 128. Blue dots represent particles, while empty sites are shown as gray dots.  
  In the case of the LG on a square lattice, the interface between the high-density and low-density regions  can be aligned with either axis of the lattice with equal probability; here we report  an instance in which it is aligned with  the vertical direction. In the driven cases, both for the IDLG and the RDLG, the driving occurs here along the horizontal direction and the interface between the high-density and low-density regions is parallel to the driving direction. The temperature $T=2.0$  is below the critical temperature in all the three cases.}
 \label{fig:snap}
\end{figure*}

The equilibrium Lattice Gas (LG) dynamics consists of jump attempts of randomly chosen particles to one of its unoccupied neighbouring sites with the Metropolis rate $w(\Delta \mathcal{H})= \min \{1, e^{-\beta \Delta \mathcal{H}} \},$ where $\beta = 1/T$ is the inverse temperature and $\Delta \mathcal{H}$ is the change in energy due to the proposed jump; see Fig. \ref{fig:cartoon_dlg} for a schematic representation.  The dynamics conserves the total number of particles in the system. The choice of the rate function $w$ ensures that the dynamics satisfies detailed balance and therefore the system eventually reaches the equilibrium state characterized by the usual Gibbs measure $P(C) \propto e^{-\beta \mathcal{H}(C)}.$
In the thermodynamic limit, the system undergoes a continuous phase transition at a critical temperature $T_c^\text{LG},$ from a disordered state to a phase-separated one where the particles cluster together: Fig. \ref{fig:snap} (left panel) shows a typical low-temperature configuration of the LG which shows the presence of a large cluster. The critical behavior characterizing this transition belongs to the Ising universality class and in $d=2$ all the equilibrium critical exponents are known exactly \cite{Baxter}.

The Driven Lattice Gas (DLG) is a generalization of the LG in which an additional nonconservative field $E$ is added along one axis of the lattice (referred to as the $\|$ direction, the one in which the lattice has length $L_\|$). This field biases the jump rates as  $w(\beta (\Delta \mathcal{H}+El))$ with $l=-1,0,1$ for jumps along, transverse or opposite to the field, respectively, as schematically represented in Fig. \ref{fig:cartoon_dlg}. Note that this dynamics describes a nonequilibrium system carrying a current of particles in the stationary state, only when the boundary condition is assumed to be periodic along the field direction. For simplicity, henceforth we consider the limiting case  $E \to \infty$, referred to as the IDLG (the `I' stands for infinite external field), in which the jumps along (opposite to) the field are always accepted (rejected). 

In the thermodynamic limit the DLG also shows a ``phase'' transition at the critical temperature $T_c^{\text{DLG}}(E),$ which surprisingly increases upon increasing $E$, saturating at a finite value $T_c^{\text{IDLG}}$ \cite{DDSbook}. For $T < T_c^{\text{DLG}}(E)$ the system shows a phase-separated state where the particles cluster in a single strip aligned with the direction of the external field; a typical low-temperature configuration of the IDLG is shown  in Fig. \ref{fig:snap} (central panel).

An important variant of the DLG is the  Randomly Driven Lattice Gas (RDLG) where the field $E$ randomly changes its sign at each attempted move.  Although this dynamics breaks the detailed balance condition, no particle current flows through the system in the stationary state of the  RDLG, in contrast to the DLG. Also for the RDLG, we consider the case  $E \to \infty$.

The RDLG also undergoes a continuous transition to a phase-separated state below a critical temperature $T_c^{\text{RDLG}}$. The low-temperature stationary state of the RDLG looks similar to that of the IDLG, the interface of the formed strip being aligned with the direction of the field $E$; see the right panel in Fig. \ref{fig:snap}. 

Both the DLG and the RDLG show  remarkable properties, such as generic long-range correlations in the disordered state and strong anisotropy in space \cite{DDSbook}. 
As a consequence of this anisotropy, finite-size scaling analysis have to be performed at a fixed anisotropic aspect ratio
\bea
S_\Delta = \frac{L_{\parallel}}{L_\perp^{1+\Delta}},
\eea
where the anisotropy exponent $\Delta$  controls the degree of anisotropy in the model. 
While $\Delta=0$ for the equilibrium LG, field-theoretical studies in two spatial dimensions conclude that $\Delta=2$ for the DLG and the IDLG \cite{JS,LC}, and $\Delta \simeq 1$ for the RDLG \cite{rdlg-S}.

In the anisotropic IDLG and RDLG, the presence of the field naturally introduces the distinction between what we refer to as ``transverse'' and ``longitudinal'' observables. A transverse observable is obtained as a spatial average along the direction of the field and is thus insensitive to spatial fluctuations along the field.  Analogously, one can also define longitudinal observables by averaging along the orthogonal direction. In the equilibrium LG such a distinction is entirely arbitrary. However, we can always fix a direction in space as being the preferred one, and define ``transverse'' and ``longitudinal'' observables  with respect to it. 

The structure of the ordered state in the driven lattice gases (see Fig. \ref{fig:snap}) naturally leads to the choice of order parameters which are transverse in nature. One such typical transverse observable used to characterize the onset of order in these systems is the so-called anisotropic order parameter $m,$   which is related to the average amplitude of the first non-zero transverse Fourier mode of the spatial density of the particles. To define it precisely, let us consider a $d=2$ dimensional lattice of size  $V= L_\| \times L_\perp$ and associate, to each site $(x,y),$ a ``spin variable'' $\sigma_{xy} = 2 n_{xy}-1$  which takes values $\pm1.$ 
The relevant quantity is its Fourier transform
\bea
\tilde{\sigma}(k_{\parallel}, k_{\perp}) = \sum_{x=0}^{L_{\parallel}-1} \sum_{y=0}^{L_{\perp}-1} e^{i(k_{\parallel} x + k_{\perp} y )} \sigma_{xy} \label{eq:sigt}
\eea
where, due to the periodic boundary conditions, the allowed longitudinal and transverse momenta are
\bea
(k_\|,k_\perp) = \left(\frac {2 \pi n_\|}{L_\|},\frac {2 \pi n_\perp}{L_\perp} \right),
\eea
with integers $0 \le n_\| \le L_\|$ and  $0 \le n_\perp \le L_\perp.$ The half-filling condition on particle density implies that the total ``magnetization'' $\sum_{xy} \sigma_{xy}$ vanishes and in turn,  $\tilde \sigma(0,0)=0.$

The anisotropic order parameter $m$ is defined as the statistical average of the absolute value of the first non-zero transverse mode $\mu =\tilde \sigma \left(0,2 \pi/ L_\perp \right),$ $i.e.,$ as
\bea
m &=&  {\left \la | \mu | \right\ra}/{V}, \label{eq:mdef}
\eea
where $\la \cdot \ra$ denotes the statistical average.
In order to characterize the dynamical behavior it is also instructive to consider the temporal auto-correlation function $C_m$ of the anisotropic order parameter, $i.e.,$
\bea
 C_m(s,t) = \frac 1V \Big[\la |\mu(s) \mu(t)| \ra  - \la |\mu(s) |\ra \la |\mu(t)| \ra \Big]. \label{eq:def-auto}
\eea

An alternative observable used to detect the onset of an ordered phase in these systems is the average absolute value $O$ of the magnetization along the field direction, which was proposed and studied in Ref.~\cite{AS2002},
\bea
O &=& \frac 1{V} \sum_{y=0}^{L_\perp-1} \left \la \left| \sum_{x=0}^{L_{||}-1} \sigma_{xy} \right| \right\ra. 
\eea
 Both $O$ and $m$  are expected to be non-zero  in the ordered phase, where the particles cluster together to form a single strip aligned along the direction of the drive, although their stationary values are expected to be different. On the other hand, both $m$ and $O$ vanish in the disordered phase in the thermodynamic limit thus providing two alternative definitions of an order parameter.

The alternative order parameter $O$ is also a transverse observable, as it can be expressed as a sum of transverse modes,
\bea
O &=& \frac 1{V^2} \sum_{y=0}^{L_\perp-1}\left \la \left| \sum_{x=0}^{L_\parallel-1} \sum_{k_{\parallel},k_\perp} e^{-i(k_{\parallel} x + k_{\perp} y )}\tilde{\sigma}(k_{\parallel}, k_{\perp}) \right| \right\ra \cr
&=& \frac 1{V} \left \la \left| \sum_{n_\perp=1}^{L_\perp-1} \tilde{\sigma}\left(0, \frac{2 \pi n_\perp}{L_\perp}\right) \right| \right\ra. \label{eq:OP}
\eea
The last equality follows from the condition $\tilde \sigma(0,0)=0$ and the fact that the expectation value on the first line does not depend on $y.$

The critical behavior of the three lattice gas models mentioned above belongs to three distinct universality classes. In  Sec. \ref{sec:ph-th} below we will briefly mention the different effective field theories which describe the critical behaviors of these models. Next, we will discuss  an effective Gaussian (non-interacting) theory and some aspects of the behavior of all the conserved lattice gases which turn out to be described by it \cite{prl-dlg}.

\section{Mesoscopic description: Field Theoretical approach} \label{sec:ph-th}

The critical behavior of the lattice gas models can be understood based on effective, mesoscopic field-theoretical descriptions of their dynamics \cite{TauberBook}. Near criticality, the evolution of the coarse-grained local spin density $\phi(x,t)$ is expected to be governed by a Langevin equation which takes into account the relevant interactions specific to each universality class. We briefly recall some basic facts about the effective theories corresponding to the lattice gases discussed here. 

{\it Equilibrium Lattice Gas}: The phase transition in the equilibrium lattice gas belongs to the Ising universality class which is characterized by the standard $\phi^4$ theory \cite{Mussardo}. Its dynamics, in the case of conserved order parameter $\phi$, is described by the corresponding Langevin equation (known as Model B) \cite{HalperinHohenberg1977, TauberBook},
\bea
\partial_t\phi = \alpha [(\tau - \nabla^2) \nabla^2 \phi ] + u \nabla^2 \phi^3 - \nabla \cdot \xi \label{eq:LG-phi}
\eea
where $\tau$ measures the distance from the critical point, $u >0$ is the interaction strength, $\alpha$ is a positive constant and $\xi$ is a while noise with $\la \xi_i(x,t) \xi_j(x',t') \ra \propto \delta_{ij} \delta^d(x-x') \delta(t-t').$
The resulting critical behavior and exponents are known exactly in $d=2$ \cite{Baxter} while the upper critical spatial dimensionality $d_c$ is 4. 

{\it Driven Lattice Gas}: The mesoscopic description of the driven lattice gas was developed independently by Janssen and Schmittmann \cite{JS} and Leung and Cardy \cite{LC}. This theory, henceforth referred to as JSLC theory, differs from the LG in two respects: the external field introduces a new interaction term and induces strong anisotropy. Correspondingly, the Langevin equation describing the evolution of the coarse-grained spin density in the near-critical DLG (and IDLG) is given by
\bea
\partial_t \phi &=& \alpha [(\tau - \nabla_\perp^2)\nabla_\perp^2 \phi + \tau_\parallel \nabla_\parallel^2 \phi + {\cal E}\nabla_\parallel \phi^2  ] \cr
&& + u \nabla_\perp^2 \phi^3 - \nabla \cdot \xi,  \label{eq:IDLG-phi}
\eea
where ${\cal E}$ represents the coarse-grained driving field, while $\nabla_\perp$ and $\nabla_\|$ denote spatial derivatives orthogonal and parallel to the driving directions, respectively. The presence of an additional relevant interaction term ${\cal E}\nabla_\parallel \phi^2$ and spatial anisotropy cause the critical behavior to change compared to that of the Ising universality class. 
Also in this case, the critical exponents are known exactly in all spatial dimensions $d \ge 2$  up to the upper critical dimension $d_c=5;$ of primary importance for the purpose of the present study is the anisotropy exponent $\Delta=2$ in two spatial dimensions which shall intervene in the following analysis.
The specific form of the interaction term causes the behavior of $\phi$ at vanishing parallel wavevector $k_\|=0$ to be effectively described by a non-interacting theory and therefore its transverse fluctuations are expected to be described by a simple Gaussian theory discussed below \cite{Caracciolo1}.

{\it Randomly Driven Lattice Gas}: The Langevin equation takes a different form compared to Eq. \eqref{eq:IDLG-phi} when the driving field changes sign randomly, \ie, in the case of the RDLG; the particle current is no longer relevant, but anisotropy continues to be a significant factor, resulting in the effective equation
\bea
\partial_t \phi &=& \alpha [(\tau - \nabla_\perp^2)\nabla_\perp^2 \phi + \tau_\parallel \nabla_\parallel^2 \phi] + u \nabla_\perp^2 \phi^3 - \nabla \cdot \xi. \cr
&& \label{eq:RDLG-phi}
\eea
In turn, this results in yet another universality class, different from both LG and DLG; the critical exponents are known in terms of a series expansion around the upper critical dimension $d_c =3$ \cite{rdlg-S,rdlg2}.  In addition, the anisotropy exponent $\Delta \simeq 1$ differs from that of the DLG in $d=2.$

 \begin{table}[t]
\begin{tabular}{c|c|c|c}
\hline
 & ~~~JSLC~~~ & ~~~RDLG~~~ & ~~LG~~ \cr
\hline\hline
$d_c$ & 5 & 3 & 4 \cr
\hline
$\Delta$ & 2 & 0.992 & 0 \cr
$\beta$ & 1/2 & 0.315 & 1/8 \cr
$\nu$ & 1/2 & 0.626 & 1 \cr
$\eta$ & 0 & 0.016 & 1/4 \cr
$z$ & 4 & 3.984 & 15/4 \cr
\hline\hline
\end{tabular}
  \caption{Critical exponents in $d=2$ for the JSLC \cite{JS,LC}, RDLG \cite{rdlg-S}, and LG \cite{Mussardo}. The values listed for the JSLC and the RDLG refer to the transverse exponents; those of the JSLC and LG are exact, while the ones of the RDLG are obtained approximately from a series expansion.}
  \label{tab:d2} 
 \end{table}

{\it Gaussian effective theory}: The Gaussian or non-interacting theory describes a fluctuating field in the absence of non-linear interactions. The corresponding Langevin equation for a system with locally conserved field can be obtained by setting $u=0$ in Eq. \eqref{eq:LG-phi},
\bea
\partial_t\phi = \alpha (\tau - \nabla^2) \nabla^2 \phi  - \nabla \cdot \xi.  \label{eq:G-phi}
\eea
Irrespective of the fact that the phase transitions in the three different models, namely LG, DLG and RDLG belong to three different universality classes, the short-time dynamical behaviors of certain transverse observables, after a quench to the critical point,  turn out to be very similar in all these models. In fact, as discussed in Ref.~\cite{prl-dlg},  transverse modes in all the lattice gas models show a behavior at short times  which is consistent with a free theory, $i.e.,$ the distribution of transverse modes is effectively Gaussian.  Some of the results of this section have already been briefly anticipated in Ref.~\cite{prl-dlg}; in the following we also provide additional details of that analysis.

In particular, our objective is to determine the temporal behavior of the order parameters introduced in Sec. \ref{sec:models}
for a model system which is described by an effective Gaussian theory. In order to do so we need to look at the time evolution of the transverse modes $\tilde{\sigma}_k$ (defined in Eq. \eqref{eq:sigt}) which is obtained by taking the Fourier transform of the Langevin equation  \eqref{eq:G-phi}. However, since we are interested in lattice models, the spatial gradients in that equation 
have to be interpreted as being defined on a lattice.  Consequently, the  amplitude $\tilde{\sigma}_k$ of the transverse mode $k=(0,k_\perp)$ evolves according to,
\beq
\frac{d}{dt}\tilde{\sigma}_k(t)= - \gamma_k \tilde{\sigma}_k(t) + i\hat k\, \tilde \eta_k(t) \label{eq:g_evol}
\eeq
where $\hat k = 2 \sin (k/2)$ is the lattice momentum and  
\bea
\gamma_k= \al  (\tau + \hat k^2) \hat k^2. \label{eq:gammak}
\eea
As mentioned above, $\al $ is a coarse-grained diffusion constant, possibly depending on the lattice parameters and $\tau$ measures the distance from the critical point.  Additionally, $\tilde \eta$ is a  white noise in  momentum space,  obtained by taking the Fourier transform of the noise in real space, and is also delta correlated, with  
\bea
\la \tilde \eta_k(t) \tilde \eta_{k^\prime}(t^\prime)\ra  = 2 \al  T_\eta L_{\parallel} L_\perp ~ \delta (k+k^\prime)\delta (t-t^\prime), 
\eea
where the normalization factor $T_\eta$  signifies an ``effective temperature'' associated with the noise in terms of which the fluctuation-dissipation theorem \cite{Kubo} is effectively satisfied when looking at correlations and response functions of the transverse fluctuations. Note that the noise strength in  momentum space is proportional to the volume of the lattice because of the discrete nature of the allowed momenta.

Let us consider the case in which  the system is initially in a disordered configuration corresponding to a high temperature, so that $\tilde{\sigma}_k(t=0)=0$ for all transverse modes $k.$ For this initial condition Eq. \eqref{eq:g_evol} has the solution
\beq
\tilde{\sigma}_k(t)=i \hat k \int_0^t \id s~ \eta_k (s) e^{-\gamma_k(t-s)}.  \label{eq:gauss_soln}
\eeq
This leads to a Gaussian behavior, $i.e.,$  the $k$-{th} Fourier mode has a Gaussian probability distribution $P$ at any time $t,$
\bea
P[\tilde{\sigma}_k(t)] = N_k(t) \exp {\left[-\frac{|\tilde{\sigma}_k(t)|^2}{L_{\parallel} L_{\perp}  G_{\perp}(t,k)} \right]}, \label{eq:Psig}
\eea
where $G_{\perp}(t,k)$ is the transverse propagator,
\bea
G_{\perp}(t,k) = \frac{1}{L_\parallel L_\perp} \la |\tilde \sigma_k(t) |^2 \ra 
\eea
and $N_k(t) = [\pi L_\parallel L_\perp \tilde G_\perp(t,k)]^{-1}$ is the normalization. 
The transverse propagator is easily computed from Eq.~\eqref{eq:gauss_soln},
\bea
 G_\perp(t,k) &=& \al T_\eta \frac{\hat k^2}{\gamma_k} (1- e^{-2\gamma_k t}). \label{eq:Gperp}
\eea
The anisotropic order parameter $m(t)$ (defined in Eq.~\eqref{eq:mdef}) can be calculated easily from Eqs.~\eqref{eq:Gperp} and \eqref{eq:Psig}, 
\begin{eqnarray}
m(t) &=& \frac{2 \pi N_{k_1}(t)}{L_{\parallel} L_{\perp}}  \int_{0}^{\infty} \id r ~r^2 \exp \left[ - \frac{ r^2}{L_{\parallel} L_{\perp}  G_{\perp}(t,k_1) } \right]  \cr
&=&  \sqrt{\frac \pi 4 \frac{G_\perp(t,k_1)}{L_{\parallel} L_{\perp}}} \label{eq:mfull}
\end{eqnarray}
where  $k_1 \equiv 2\pi/L_\perp$  indicates the first non-zero mode allowed in the transverse direction. 

 We are particularly interested in the dynamical behavior of $m(t)$ in the short-time regime, $i.e.,$ immediately after the critical quench, when the system is far from reaching its stationary state. In this regime, one can expand the exponential in Eq. \eqref{eq:Gperp} and keep only the linear term in $t,$ finding,  for any $\tau,$  
\bea
G_{\perp} \left( t,k_1 \right)  =  2 \al  T_\eta {\hat k_1}^2~ t + {\cal O}(t^2). 
\eea
In the thermodynamic limit, $i.e.,$ for large $L_\perp,$ we have ${\hat k}_1 = 2 \sin (k_1/2) \simeq 2 \pi /L_\perp.$  To the leading order in $t,$ then, 
 Eq.~\eqref{eq:mfull} implies,
\begin{equation}
\label{eq:mt}
m(t) \approx \sqrt{ 2 \pi^3 \al T_\eta \frac t{L_{\parallel} L_{\perp}^3}},
\end{equation}
for $t \ll L_\perp^4.$ This spells a clearer meaning to the term short-time regime: this behavior is expected to hold up to a time which is much shorter than the time-scale set by the system size.

At longer times, instead,  $m(t)$ approaches a stationary value $m_S$ which can also be obtained from Eq.~\eqref{eq:mfull}. In particular, at the critical point $\tau=0,$
\bea
m_S \equiv \lim_{t \to \infty} m(t) = \sqrt{\frac{T_\eta}{16 \pi}\frac{L_\perp}{L_\|}}, \label{eq:m-st}
\eea
which depends only on the isotropic aspect ratio $L_\perp/L_\|.$

In order to predict the behavior of the order parameter $O,$  defined in Eq.~\eqref{eq:OP}, we first note that as each single mode $\tilde \sigma_k$ (see Eq. \eqref{eq:g_evol}) is a stochastic variable with a Gaussian distribution, their sum in Eq.~ \eqref{eq:OP} also has a Gaussian distribution. Accordingly, 
\bea
O(t) = \sqrt{\frac \pi 4 \frac{D(t)}{L_\perp L_{||}}}, 
\eea
where $D(t)$ is the sum of the transverse propagators of the modes appearing in Eq.~\eqref{eq:OP}, $i.e.,$
\bea
D(t) = \sum_{n_\perp=1}^{L_\perp -1} G_\perp \left(t, \frac{2 \pi n_\perp}{L_\perp} \right). \label{eq:Dt}
\eea
For sufficiently large $L_\perp$ one can take the continuum limit of this expression and the sum over $n_\perp$ is replaced by a momentum integral; at the critical point $\tau=0,$ one then finds,
\bea
D(t) & \simeq &  T_\eta L_\perp \int_{-\pi}^\pi \frac{\id k}{2 \pi} \frac {1- e^{-2\al  t k^4}}{k^2} \cr
&=& \frac{T_\eta L_\perp}{\pi^2} \left\{ (2 \al  t)^{1/4} \pi \left[\Gamma\left(3/4 \right)-\Gamma\left(3/4,2 \al  t \pi^4 \right) \right] \right. \cr
&& \qquad \quad \left. + e^{-2 \al  t \pi^4}  -1 \right \}, 
\eea
where $\Gamma(x)$ is the Gamma function and $\Gamma(x,s)$ is the incomplete Gamma function; see, \eg,  Eq.~8.2.2 in Ref.~\cite{math-book}.
In particular, for large enough $\al t \gg 1,$ 
\bea
D(t) &\simeq&  \frac {T_\eta L_\perp}{\pi} \Gamma\left( 3/4 \right) (2\al  t)^{1/4}.    
\eea
Accordingly, from Eq. \eqref{eq:Dt},  $O$  grows, in this intermediate time regime, as
\bea
O(t) \simeq \frac{t^{1/8}}{2L_{\parallel}^{1/2}} \sqrt{T_\eta(2 \al )^{1/4} \Gamma (3/4) }, \label{eq:Ot}
\eea 
$i.e.,$  $O(t) \sim t^{1/8}$ upon increasing $t.$
We emphasize here that the limits $L_\perp \to \infty$ and $t \to \infty$ do not commute. To obtain the stationary value $O_S$ of $O(t),$ one can perform a direct summation in Eq.~\eqref{eq:Dt} with $G_\perp$ given by Eq. \eqref{eq:Gperp}, and get,
\bea
\lim_{t \to \infty} D(t) = T_\eta \sum_{n_\perp=1}^{L_\perp -1} \frac 1{2 \sin(\pi n_\perp/L_\perp)} = \frac {T_\eta}{12} (L_\perp^2 -1).\n
\eea
Accordingly, from Eq. \eqref{eq:Dt}, assuming $L_\perp \gg 1,$
\bea
O_S \equiv \lim_{t \to \infty} O(t) = \sqrt{\frac{T_\eta \pi}{48}\frac{L_\perp}{L_\|}}. \label{eq:O-st}
\eea

One comment is in order here.  As we will show in the next Section, the short-time behavior of the order parameters $m$ and $O$ predicted on the basis of the effective Gaussian theory (in Eqs.~\eqref{eq:mt} and \eqref{eq:Ot}) holds in driven lattices gases irrespective of the system size and of any specific geometrical aspect ratio of the lattices \cite{prl-dlg}. On the other hand, the stationary state, reached at larger times, is different for the various lattice gases and it is only for the specific case of IDLG that the JSLC theory predicts a Gaussian behavior of transverse modes, also in the stationary state. Consequently, the behaviors of $m$ and $O,$ in the stationary state,  as predicted by the Gaussian theory (in Eqs. \eqref{eq:m-st} and \eqref{eq:O-st}, respectively) are expected to hold only for the IDLG  assuming the appropriate anisotropic scaling.

The auto-correlation $C_m$ of the order parameter $m,$ defined in Eq. \eqref{eq:def-auto}, can also be easily calculated within the Gaussian model discussed here. The joint distribution of $\tilde \sigma_k(s)$ and $\tilde \sigma_k(t)$ following from Eq. \eqref{eq:gauss_soln} is nothing but a multi-variate Gaussian distribution \cite{multi-G},
\bea
 P[\tilde{\sigma}_k(s), && \, \tilde{\sigma}_k(t)] \, =\frac 1{4 \pi^2 {\cal D}} \exp \left\{- \frac 1{2{\cal D}} \left[\lambda_{tt} |\tilde \sigma_k(s)|^2  + \right. \right.\cr
&&\left.   \phantom{\frac{1}{4}} \left. \lambda_{ss} |\tilde \sigma_k(t)|^2 - 2\lambda_{st}{\text {Re}\,}(\tilde \sigma_k(s)  \tilde \sigma^*_k(t)) \right ] \right \}  
\eea
where $*$ denotes the complex conjugate, and
\bea
\lambda_{t_1t_2} &=& \big \la [\text{Re}~\tilde{\sigma}_k(t_1)][\text{Re}~\tilde{\sigma}_k(t_2)]  \big \ra\cr
&=&\frac{\al  T_\eta V}2 \frac{\hat k^2}{\gamma_k} e^{-\gamma_k (t_2-t_1)}(1-e^{- 2 \gamma_k t_1}), \label{eq:abd}
\eea
with ${\cal D} = \lambda_{ss} \lambda_{tt}- \lambda_{st}^2 > 0.$
To obtain the auto-correlation $C_m(s,t)$ of the lowest mode $\mu$ with $k=k_1,$ we need to compute a double spherical integral,
\bea
 \la| \mu(s) \mu(t)| \ra &=& \frac 1{4 \pi^2 {\cal D}} \int_0^\infty \id r_1 ~ \id r_2 \int_0^{2 \pi} \id \theta_1~  \id \theta_2 ~r_1 ^2  r_2 ^2 ~ \times \n \\ [0.25 em]
\exp \Big\{&-& \frac 1{2{\cal D}} [\lambda_{tt} r_1^2 + \lambda_{ss} r_2^2 - 2 \lambda_{st}~ r_1 r_2 \cos(\theta_2-\theta_1) ] \Big \} \n \\  [0.5em] 
&=& \sqrt{\lambda_{tt} \lambda_{ss}}~ \left[2 E\left(y \right) - \left(1- y \right) K\left(y \right) \right ] \label{eq:msmt}
\eea
where $y=\lambda_{st}^2/(\lambda_{tt} \lambda_{ss}).$  Here $K(x)$ and $E(x)$ are the Legendre's complete Elliptic integrals of the first and second kind, respectively; see Sec.~19.2 in Ref.~\cite{math-book}. 

In the short-time regime where $s < t \ll \gamma_k^{-1},$  one has $\lambda_{ss}\lambda_{tt} \simeq (\al  T_\eta V)^2 {\hat k}^4 st$ and $\lambda_{st}^2/(\lambda_{tt} \lambda_{ss}) \simeq s/t.$ Moreover, for small $x,$
\bea
K(x) &=& \frac \pi 2 + \frac {\pi x}8 + {\cal O}(x^2), \n \\[0.25em]
E(x) &=& \frac \pi 2 - \frac {\pi x}8 + {\cal O}(x^2). \label{eq:EE}
\eea

Combining Eq.~\eqref{eq:msmt} with Eq.~\eqref{eq:EE} and using Eq.~\eqref{eq:mfull} yields the connected correlation function (defined in Eq.~\eqref{eq:def-auto});  to the leading order in $s/t$,
\bea
C_m(s,t) = \al  T_\eta {\hat k}^2 \frac \pi 8~ t \left({\frac st} \right)^{3/2}. \label{eq:Cmst}
\eea
This behavior is expected to hold in the short-time regime, i.e., for $s/t \lesssim 1.$

A useful indicator of deviation from the Gaussian behavior is the so-called Binder cumulant $g$ \cite{Binder}. Its appropriate definition for systems with conserved order parameter  has been proposed in Ref.~\cite{Caracciolo2},
\bea
g = 2- \frac{\la |\mu|^4 \ra}{\la |\mu|^2 \ra^2}, \label{eq:binder} 
\eea
where $\mu$ is defined before Eq. \eqref{eq:mdef}.  For a Gaussian field, $\la |\mu|^4 \ra  = 2 \la |\mu|^2 \ra^2 =2 V^2  G_{\perp}^2 $ and thus the Binder cumulant vanishes. Its possible finite value is therefore a good measure of the deviation from a Gaussian behavior. 


In the following Secs. \ref{sec:short-t} and \ref{sec:long-t} we compare the predictions of the Gaussian theory with the results of numerical simulations in the three different lattice gas models, both in the short-time regime and in the stationary state.

\begin{figure}[t]
 \centering
 \includegraphics[width=8.8 cm]{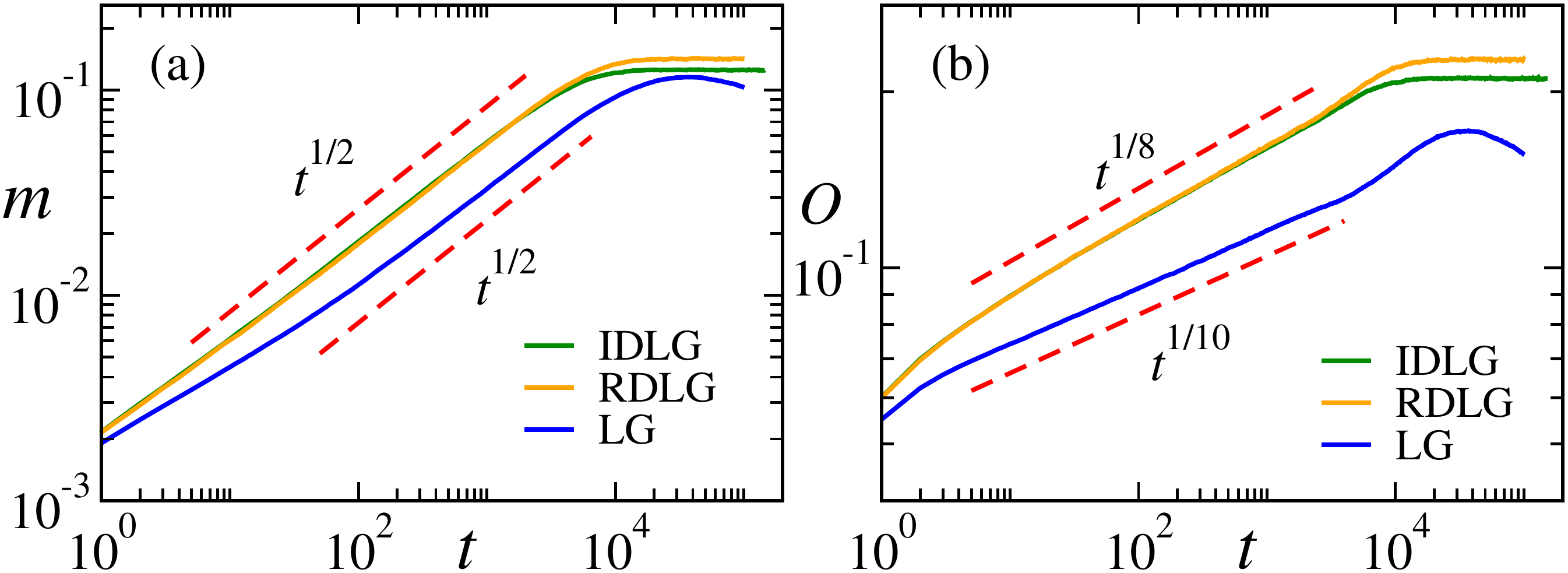}
 \caption{Short-time dynamical behavior of the order parameters $m$ (a) and $O$ (b) at $T=T_c$ in all the lattice gas models: RDLG (topmost curves in both panels), IDLG (middle curves) and LG (lowest curves), on a  $L_\| \times L_\perp = 128 \times 32$ lattice. The anisotropic order parameter $m$ grows as $t^{1/2}$ as a function of time $t$ in all cases in panel (a).  In panel (b), $O$ grows as $t^{1/8}$ upon increasing $t$ both for the IDLG and the RDLG, while it grows as $t^{1/10}$ for the equilibrium LG.}
 \label{fig:growth}
\end{figure}

\section{The short-time regime}\label{sec:short-t}

We perform Monte Carlo simulations to determine the dynamical behavior of the order parameters $m$ and $O$ and their auto-correlations in all the three lattice gas models introduced above, namely, LG, IDLG, and RDLG. The simulations are done on two-dimensional rectangular lattices of size $L_{\|} \times L_\perp$ where $\|$ and $\perp$ denote the directions parallel and transverse to the driving field in IDLG and RDLG, and arbitrary directions in LG. Periodic boundary conditions are assumed in both the spatial directions. Each Monte Carlo step, which sets the unit of time, consists of $V=L_{\|} L_\perp$ attempted jumps. 

\begin{figure*}[t]
 \centering
\includegraphics[width=5.8 cm]{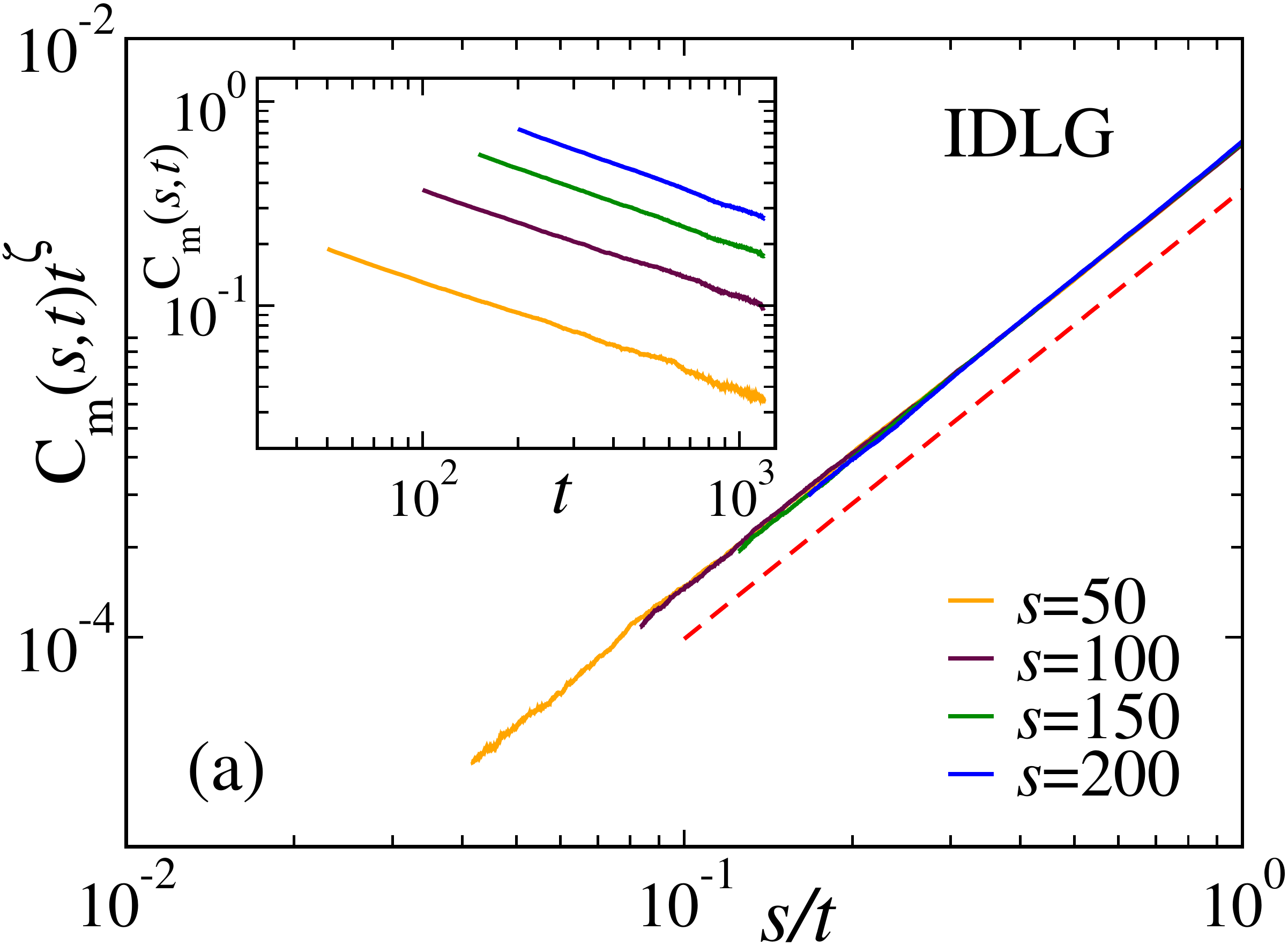}\hspace*{0.02 cm}
 \includegraphics[width=5.8 cm]{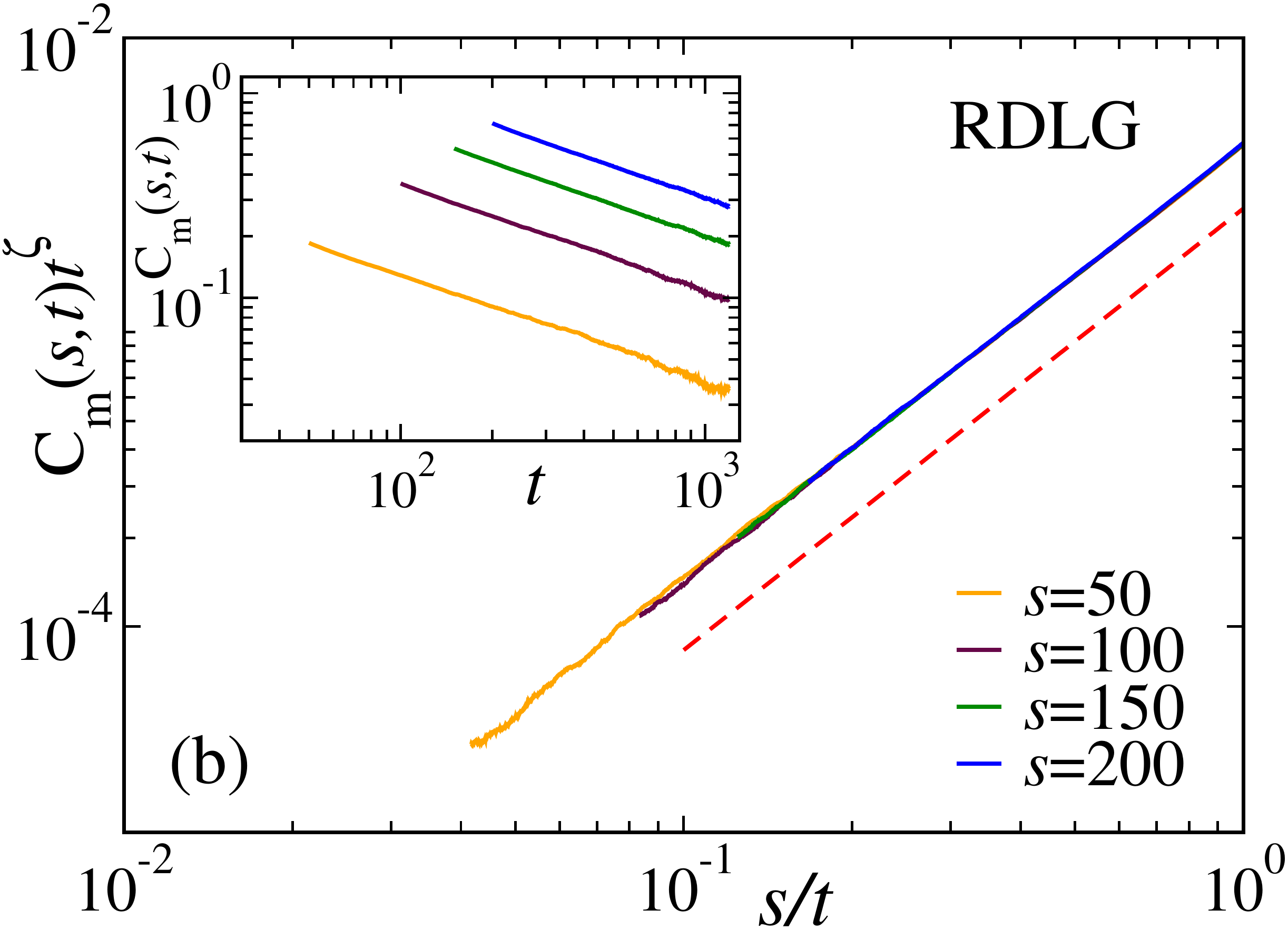}\hspace*{0.02 cm}
\includegraphics[width=5.8 cm]{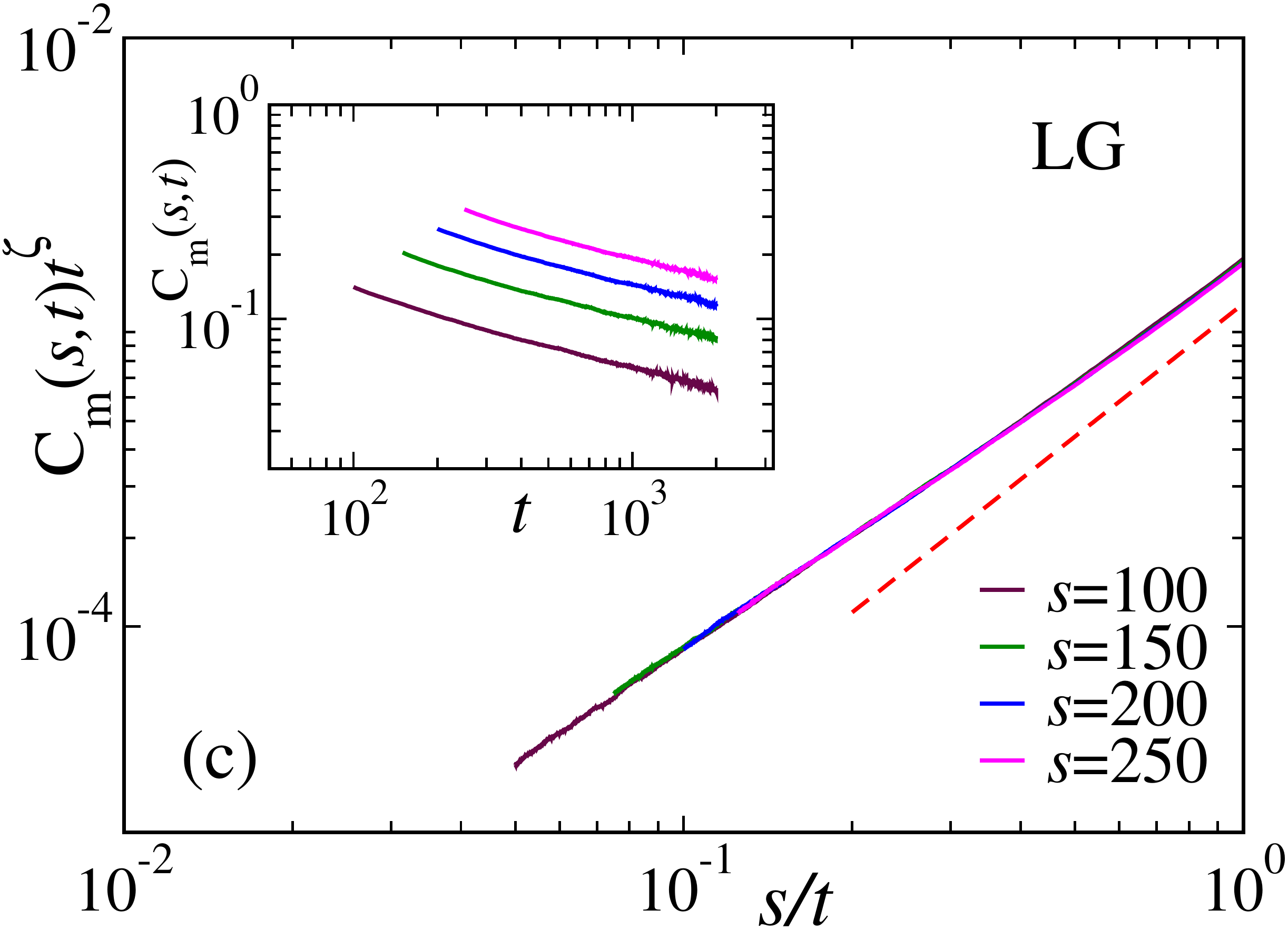}
 \caption{Plot of $C_m (t,s) t^\zeta$ as a function of $s/t$ for (a) IDLG,  (b) RDLG and (c) LG  and a set of values of $s.$ The best collapse of the curves is attained with $\zeta=0.96(2)$ for IDLG, $\zeta=0.96(2)$ for RDLG and $\zeta=0.95(1)$ for LG. The dashed red lines correspond to $(s/t)^{3/2}$ and the insets show the unscaled data in all three cases.  The system size is $L_\| \times L_\perp = 128 \times 32$ in all panels. }
 \label{fig:Cm}
\end{figure*}

In each case, the system is prepared initially in a disordered configuration corresponding to the stationary state at $T \to \infty$ in which both the order parameters $m$ and $O$ vanish. The time evolution is studied at the critical temperature $T_c,$ which is different for the three models with  $T_c^{\text{LG}}= 2.269$ \cite{Baxter},  $T_c^{\text{IDLG}}=3.20$ \cite{Tauber} and $T_c^{\text{RDLG}}=3.15$ \cite{prl-dlg}, (see also Sec.~\ref{sec:tcrdlg} below for the determination of $T_c^{\text{RDLG}}$) respectively. In  Sec. \ref{sec:binder} below we will also consider the time evolution of the Binder cumulant starting from different initial conditions and study how this affects the short-time Gaussian behavior.


\subsection{Evolution of the order parameters}

The behavior of the order parameters $m$ and $O$ agree very well  with the predictions of the Gaussian theory in Eqs. \eqref{eq:mt} and \eqref{eq:Ot}, the only exception being the case of $O$ in LG \cite{prl-dlg}. This can be seen in Fig. \ref{fig:growth} where we compare  $m(t)$ and $O(t)$ for different models  for the same system size. In the short-time regime the curves corresponding to IDLG and RDLG are almost identical with $m(t) \sim t^{1/2}$ and $O(t) \sim t^{1/8}.$ For LG, instead, $O(t) \sim t^{1/10}$
while the anisotropic order parameter  $m(t)$ still shows a $\sim t^{1/2}$ growth, consistent with a Gaussian behavior  (see Ref.~\cite{prl-dlg} for more details).     

The Gaussian theory provides a way to determine the normalization constants  $\al $ and  $T_\eta$ independently. From a fit of the  curves of $m(t)$ and $O(t)$ according to Eqs. \eqref{eq:mt} and \eqref{eq:Ot} in the short-time regime (excluding possible lattice effects for very small $t$) one can determine
the combinations $\al T_\eta  $ and $\al ^{1/4} T_\eta$ respectively. These values along with the individual estimates  of  $\al$ and $T_\eta$ obtained using them are reported in Table \ref{tab:Zl}. The values of $\al $ and $T_\eta$ for the IDLG and the RDLG are very close, consistent with their behavior as seen in Fig. \ref{fig:growth}. We have also checked that these values do not depend significantly on the system size. For the LG, instead, only $m$ follows the Gaussian prediction and we can determine the combination $\al T_\eta $ only, not the individual parameters and this estimate of $\al T_\eta$ (see lowest row on Table \ref{tab:Zl}) differs considerably from those for the driven lattice gases. 

It is interesting to note that $T_\eta/T_c$ is very close to unity for both IDLG and RDLG.
This suggests that the dynamics of the lowest transverse modes at short-times is not only ruled by an effective Gaussian model leading to a linear Langevin equation, but also that this dynamics occurs as in an equilibrium system at the same temperature as that ruling the particle transitions on the lattice transversely to the driving field.

\begin{table}[t]
\begin{tabular}{l|c|c||c|c|c}
\hline
 & $\al T_\eta  $ &  $\al ^{1/4} T_\eta  $  & $\al $ & $T_\eta$ & $T_\eta/T_c$ \cr
\hline\hline
IDLG & 0.23264 &  1.6485  & 0.0735 & 3.166 & 0.989 \cr
RDLG & 0.23374 &  1.62275  & 0.0755 & 3.096 & 0.983 \cr
LG & 0.06978 &  --  & -- & -- & -- \cr
\hline
\end{tabular}

\caption{Values of normalization factors $T_\eta$ and $\al $ as obtained from the temporal growth of $m(t)$ and $O(t)$ in the different lattice gas models. The system size $L_\| \times L_\perp$ used to determine these quantities are $1024 \times 64$ for the IDLG, $144 \times 48$ for the RDLG, and $128 \times 64$ for the LG.} \label{tab:Zl}
\end{table}


\subsection{Auto-correlation of the order parameter}

 Close to a phase transition, the temporal auto-correlation of the order parameter also typically carries the signature of the universal critical behavior \cite{Godreche}. This fact has been used in the literature to distinguish between different universality classes of driven lattice gases by studying, \eg, the particle density auto-correlation of the IDLG \cite{Tauber}. In this view, it is interesting to explore the behavior of the auto-correlation of the anisotropic order parameter $m$ for the various lattice gas models and compare it with the prediction of the Gaussian theory. 

To this end, we measure the auto-correlation $C_m(s,t)$ (defined in Eq.~\eqref{eq:def-auto}) of the order parameter $m$ in the short-time regime after a critical quench for all the three models using Monte Carlo simulations.  Figure \ref{fig:Cm} shows plots of $C_m(s,t) t^\zeta$ as a function of $s/t$ for IDLG (panel (a)), RDLG (panel (b)) and LG (panel (c)) where $\zeta$ is the exponent obtained from the best collapse of the data. In particular, we obtain, $\zeta = 0.96(2)$ for the IDLG,  $\zeta = 0.96(2)$ for the RDLG and  $\zeta = 0.95(2)$ for the LG. All these three values agree rather well with the  prediction $\zeta =1$ of the Gaussian theory, see Eq.~\eqref{eq:Cmst}. Moreover, the behaviors of the scaled curves is also consistent with the Gaussian theory in all the cases, showing a growth $\sim (s/t)^{3/2}$ upon increasing $s/t$ (dashed red lines in Fig.~\ref{fig:Cm}).  Accordingly, we conclude that it is not possible to distinguish between the different lattice gas models even on the basis of the  the auto-correlation of the anisotropic order parameter $m$ in the short-time regime.

However, it is to be noted that other two-time quantities like the density auto-correlation, which cannot be expressed as a function of the transverse modes only, can be successfully used in order to discriminate the different models even in the short-time regime, as it has been demonstrated in Ref.~\cite{Tauber}. 
This fact clearly shows that, in the presence of a local conservation law, an attentive choice of observables is necessary in order to be able to distinguish between different universality classes and that some choices turn out to be inadequate at short-times in spite of the fact that they naturally appear as being bona fide order parameters.

\subsection{Binder cumulant: dependence on the initial condition}\label{sec:binder}

The Binder cumulant $g$ is an effective measure of Gaussian behavior or deviation therefrom. Beyond its widespread applications in equilibrium statistical physics, it has also been used in the context of nonequilibrium lattice gases
in order to characterize the stationary state behavior \cite{Caracciolo2,rdlg2}. More recently, $g$ has been used to show that the dynamical behavior of the first non-trivial transverse mode is well described by a Gaussian theory up to a time which scales as $L_\perp^z$ in all the three different lattice gas models \cite{prl-dlg}.
However, the stationary value of the Binder cumulant conclusively distinguishes between these three universality classes. It is therefore natural to ask what is the origin of the observed ``super-universal'' Gaussian behavior in the short-time regime and, via the analysis of the behavior of $g$, to investigate how much of it depends on the specific choice of the initial condition, chosen to be disordered in Ref. \cite{prl-dlg}. Accordingly, in the following we explore the dynamical behavior of the Binder cumulant starting from the various initial conditions depicted schematically in Fig. \ref{fig:schem}.  In particular, we consider the following configurations:

\begin{figure}[h]
\centering
\includegraphics[width=2.8 cm]{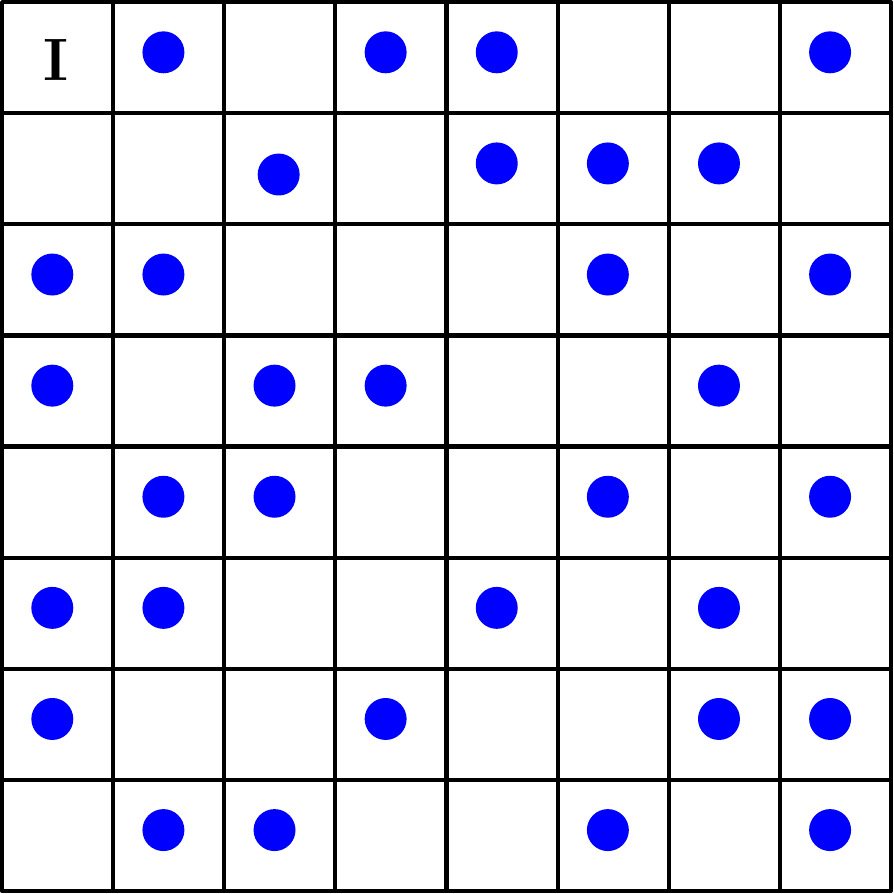}
\hspace*{0.1 cm}
\includegraphics[width=2.8 cm]{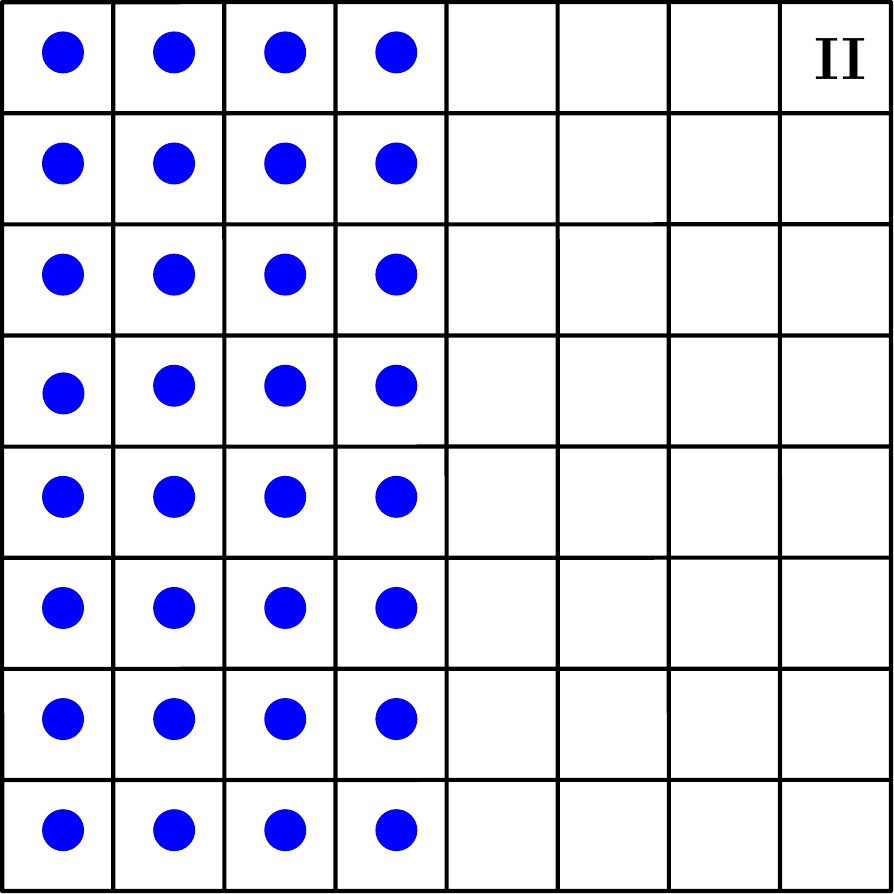} \\
\vspace*{0.2 cm}
\includegraphics[width=2.8 cm]{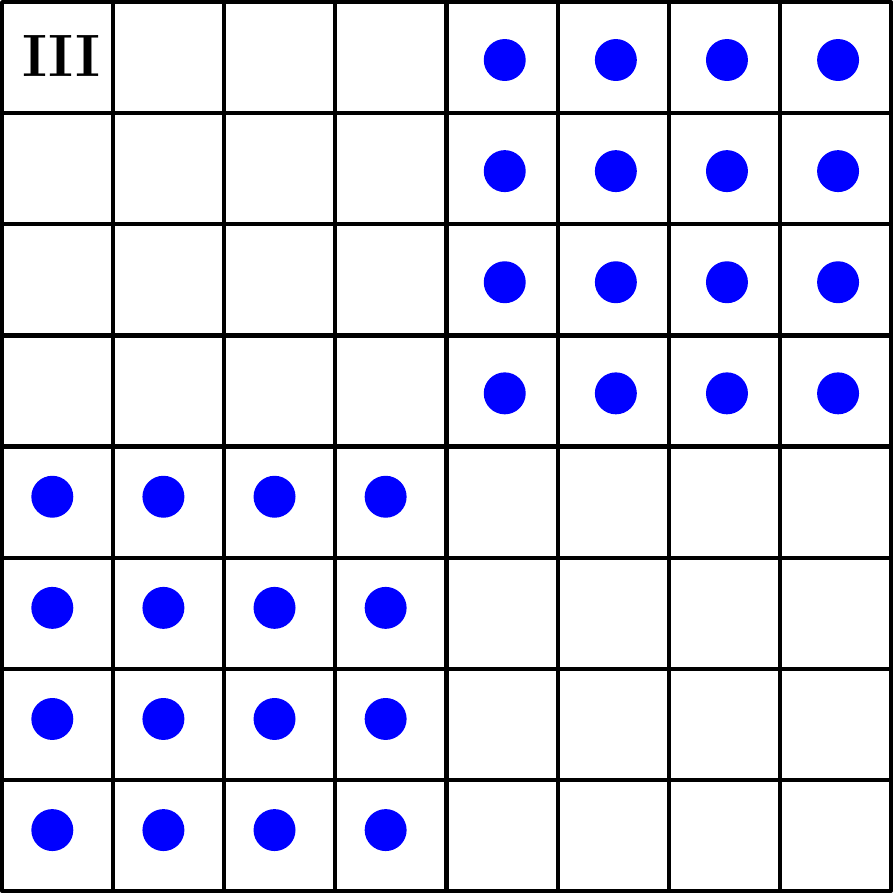}
\hspace*{0.1 cm}
 \includegraphics[width=2.8 cm]{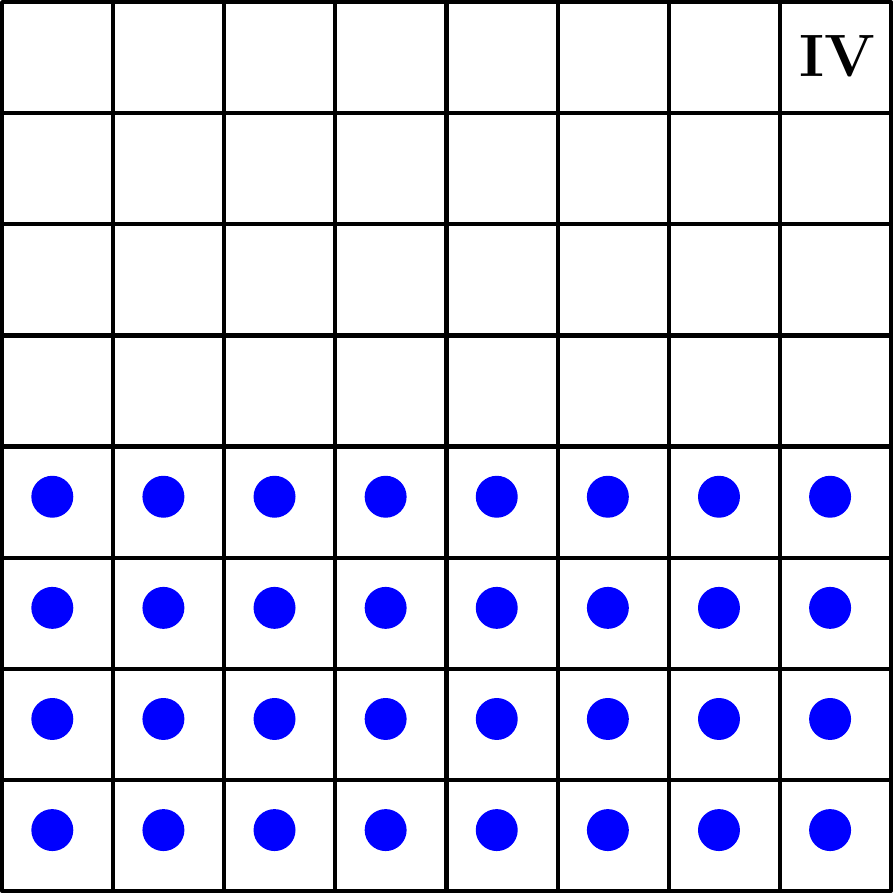}
 \caption{Schematic representations of the various initial conditions considered in Sec.~\ref{sec:binder}. The blue dots correspond to particles. In the driven systems the field acts along the horizontal direction.}
 \label{fig:schem}
\end{figure}

\begin{figure*}[t]
 \centering
\includegraphics[width=5.8 cm]{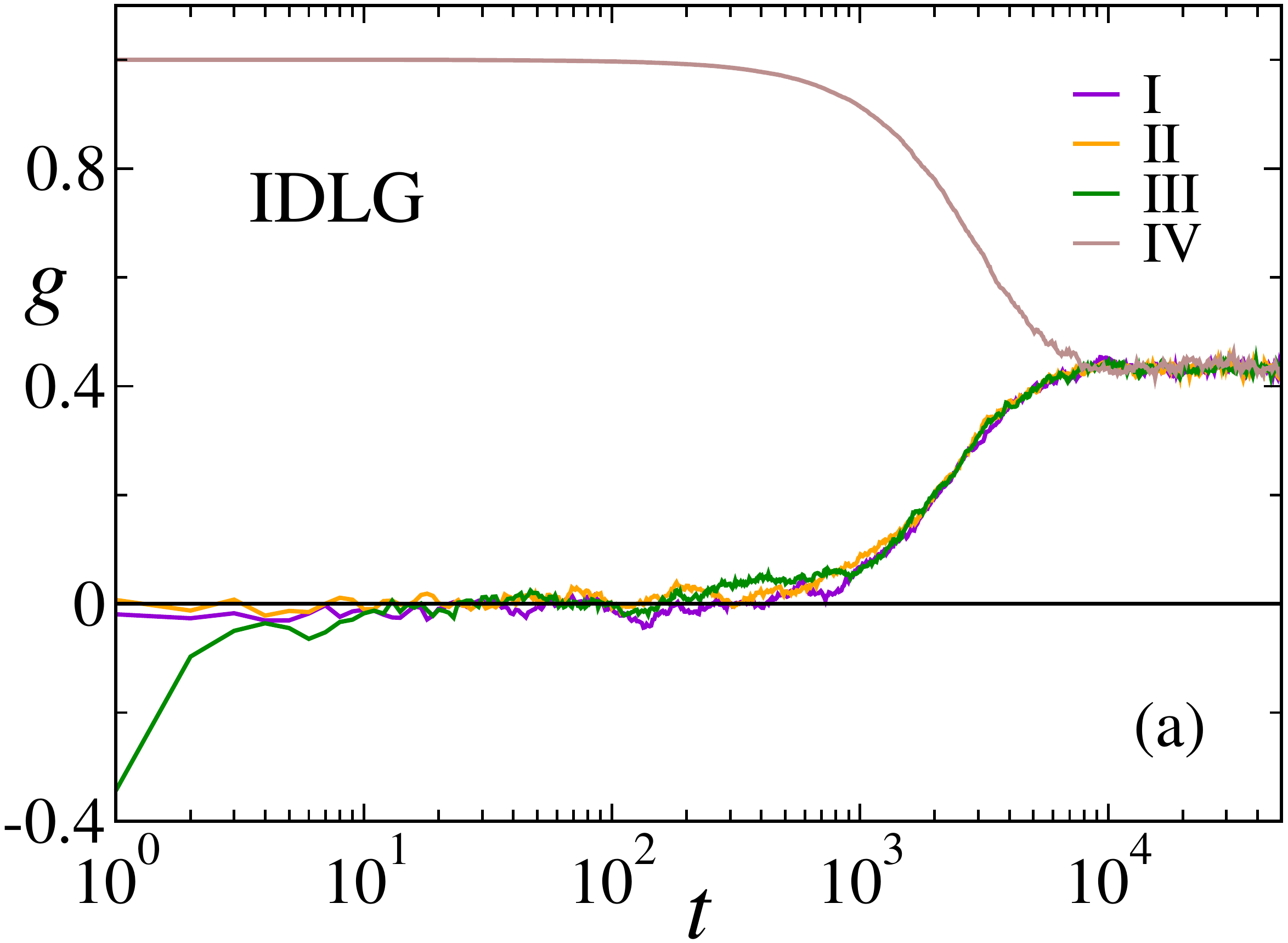}\hspace*{0.02 cm}
 \includegraphics[width=5.8 cm]{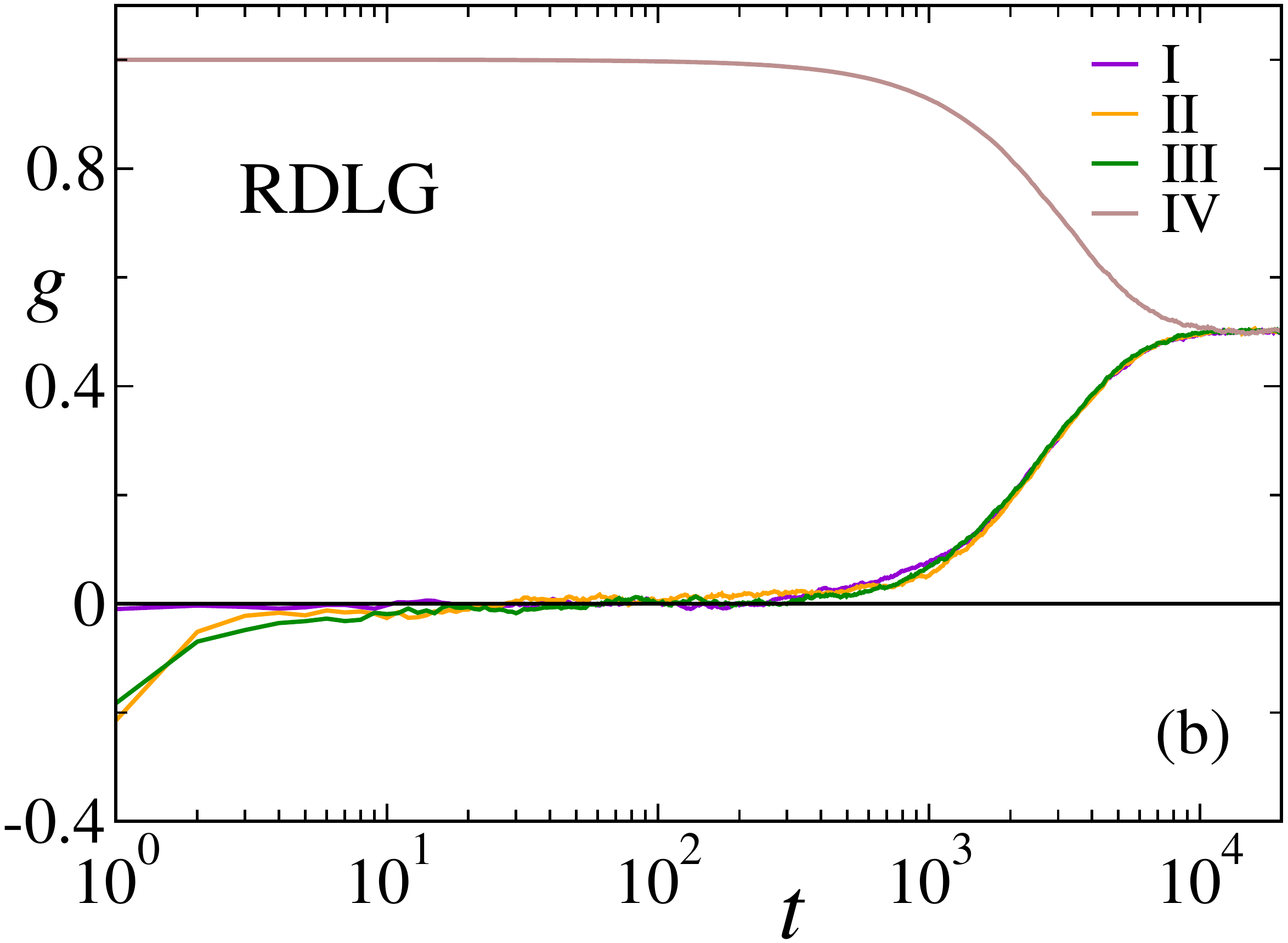}\hspace*{0.02 cm}
\includegraphics[width=5.8 cm]{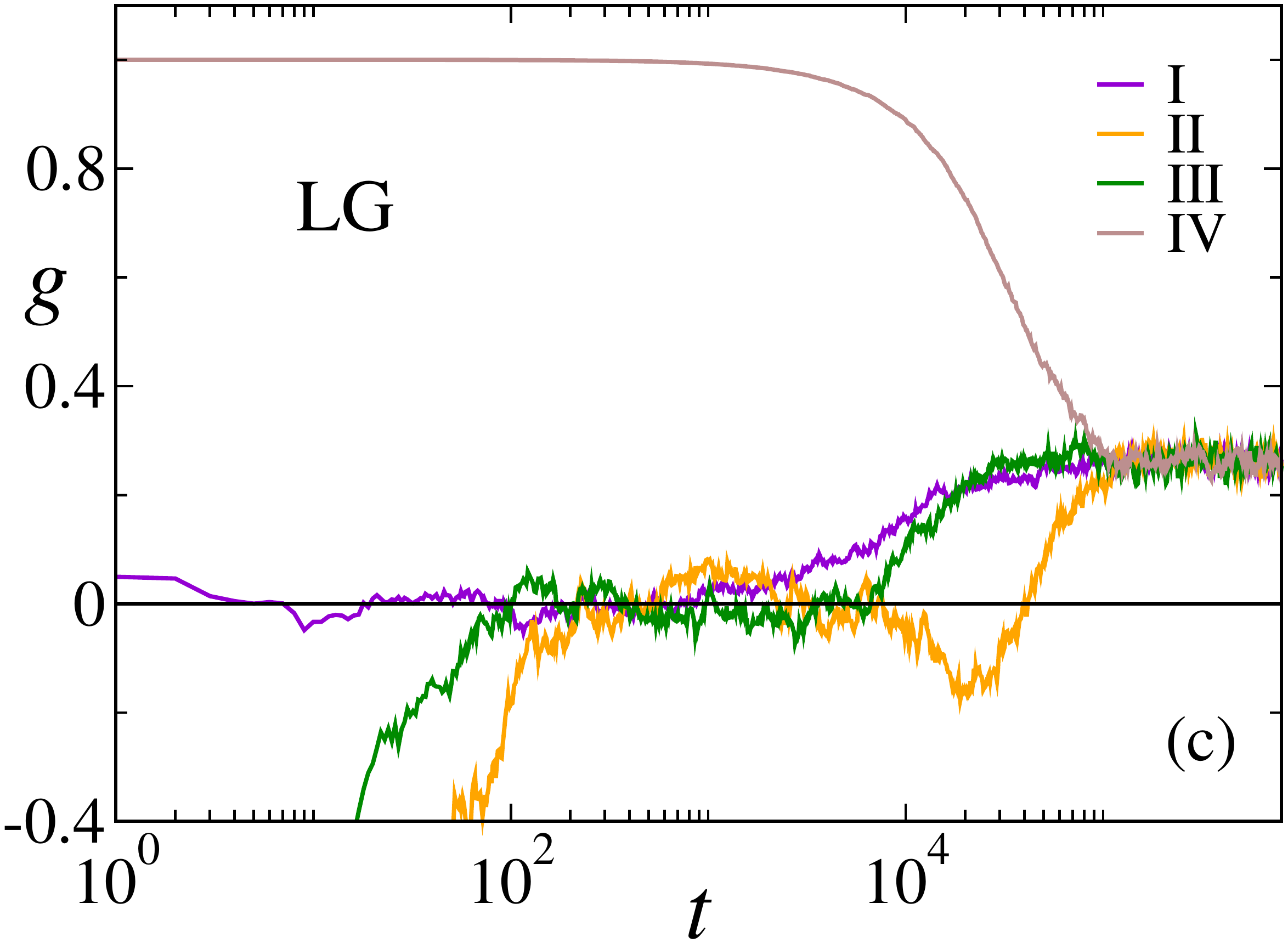} 
 \caption{Time evolution of the Binder cumulant $g$ for (a) IDLG, (b) RDLG, and (c) LG and the various initial conditions indicated by I, II, III, and IV in Fig.~\ref{fig:schem}. The numerical data have been obtained from Monte Carlo simulations on a lattice of size $L_\| \times L_\perp = 32 \times 32$ in all the cases.}
 \label{fig:init}
\end{figure*}

\begin{itemize}

\item [I.] Disordered configuration: This corresponds to a typical configuration at high temperature, as the particles are distributed randomly; we ensure that the magnetization on each row is exactly zero so that $m$ vanishes in this state. This initial condition is the one used to study the behavior of the order parameters in Sec. \ref{sec:short-t}.  


 \item [II.] Column-ordered configuration:  This initial condition resembles  a phase-separated state but  the interface is orthogonal to the direction of the field and hence the anisotropic order parameter $m$ vanishes. 
Note that this configuration corresponds to one of the two equivalent low-temperature configurations of the LG on a square lattice. 

 \item [III.]  Mixed-ordered configuration: For this initial condition the particles are arranged on the lattice in order to form a checker board pattern, the top right and bottom left sub-rectangles are the only ones being occupied. 
Also in this configuration the order parameter $m$ vanishes. 
%
 \item [IV.] Row-ordered configuration: Here we start from the phase separated state, with the interface being parallel to the direction of the drive. In the case of the LG, this is taken to be the $x$-direction, mimicking the ordered configuration in the driven cases. This configuration corresponds to a finite non-zero value of $m.$ 
 
\end{itemize}

The Binder cumulant $g$ is computed, as discussed in Sec.~\ref{sec:ph-th}, with reference to the first transverse mode, according to Eq. \eqref{eq:binder}. 

Figure \ref{fig:init}  shows plots of the time evolution of the Binder cumulant $g$ starting from these various initial configurations for all the three lattice gas models, at the corresponding critical temperatures.
Although the three initial conditions  I, II, and III all correspond to a vanishing value of the order parameter $m,$ the particle distributions in space are very different in the three cases. However, in each case, after an initial transient there is an intermediate regime where the transverse fluctuations are Gaussian, as indicated by the vanishingly small value of $g$ (see the light orange, dark green,  and purple curves in Fig. \ref{fig:init}).  This observation reinforces the idea that, at criticality, the short-time evolution  of the transverse modes of the lattice gases is indeed governed by a Gaussian dynamics as in Eq. \eqref{eq:gauss_soln}  as long as the initial configuration of the lattice is a not-ordered one, $i.e.,$ with a vanishing initial value of the order parameter.
 
Note that, the LG, in contrast with the IDLG and the RDLG, shows a more pronounced initial nonzero stretch. Also, for the LG with the column initial condition II, the onset of growth of the Binder is marked by an unexpected dip.  These features bear signature of the fact that the LG is, in some way, ``less Gaussian'' than the driven lattice gases.  
In fact, it is rather surprising that the Binder cumulant shows a vanishingly small value for a considerably long time for the LG (see Fig. \ref{fig:init}(c)), because, as it is well known, it is actually described by an interacting $\phi^4$ theory  characterized by a non-vanishing stationary value of the cumulant $g$ \cite{Mussardo}. 

In order to understand the short-time behavior of the Binder cumulant in the LG model we perform a perturbative analysis for the $\phi^4$ theory around the Gaussian fixed point. We calculate the evolution of  $g$ for a small interaction strength $u,$ as defined in Eq. \eqref{eq:LG-phi}. It turns out that the growth of $g$ is slowed down by a factor of $k^4$ compared to the non-conserved case.  Consequently, for the first transverse mode with the smallest value of $k,$ $g$ appears to be vanishingly small; see Appendix \ref{sec:pert} for the details.

The configuration IV corresponds to an ordered state, and in this case,  for all the models considered, the Binder cumulant $g$ starts from unity and monotonically decreases towards the stationary value (uppermost, light brown curves in the plots of Fig.~\ref{fig:init}). In the long-time limit, as expected, $g$ attains the same stationary value irrespective of the initial conditions, depending only on the specific model. This is clearly shown by all curves in Fig.~\ref{fig:init}.

\section{The Stationary state} \label{sec:long-t}

The stationary state of the lattice gas models bears the signatures of the specific universality class, displaying different behavior for the three different models considered in this study. In the case of the IDLG, also the stationary behavior of transverse observables is described by the Gaussian theory in the limit of large system size, as predicted by the corresponding JSLC theory \cite{JS,LC}.  However, this is not the case for LG and RDLG, the stationary properties of which are significantly different from those predicted by a Gaussian theory.
In the following we discuss the stationary behavior of the order parameters $m$ and $O$ in the driven lattice gases and compare them with the Gaussian behavior.

\subsection{Stationary values of the order parameters}

The predictions of the Gaussian theory for the stationary values of $m$ and $O$ are reported in Eqs. \eqref{eq:m-st} and \eqref{eq:O-st}, respectively. Accordingly, at the critical temperature, these transverse observables in the IDLG should reach the stationary values,
\bea
m_S = \sqrt{\frac {T_\eta} {16\pi} \frac {L_\perp}{L_{||}}}, \quad  \text{and} \;\; O_S = \sqrt{\frac {T_\eta\pi} {48} \frac {L_\perp}{L_{||}}}. \label{eq:mOS_gauss}
\eea
which depend on the geometry of the lattice only via the isotropic aspect ratio $L_\perp/ L_\|.$  

An alternative way to predict the finite-size behavior of $m_S$ and $O_S$ is to use the scaling theory which demands that, at the critical point, the order parameter vanishes as 
\bea
m_S \sim L_\perp^{-\beta / \nu}, \label{eq:mL_scaling}
\eea
upon increasing the system size $L_\perp$ \cite{Caracciolo2}. To connect the prediction of the scaling theory with that of the Gaussian theory, we need to  express the behavior of $m_S$ as a function of the isotropic aspect ratio $L_\perp/L_\|.$  In order to do so, we remember that the finite-size scaling of the driven lattice gases has to be performed at a fixed anisotropic aspect ratio $S_\Delta= L_\|/L_\perp^{1+\Delta}$ \cite{DDSbook} which, in turn, implies $L_\perp/L_\| \sim L_\perp^{-\Delta}.$ Then, from Eq. \eqref{eq:mL_scaling} we have, 
\bea
m_S \sim \left(\frac{L_\perp}{L_\|} \right)^{\beta/ (\nu \Delta)}. \label{eq:m_aspect}
\eea
$O_S$ is also expected to scale in the same way.

For the IDLG, $\Delta = 2$ and $\beta/\nu =1$ (see Table \ref{tab:d2}) and Eq. \eqref{eq:m_aspect} is  compatible with the prediction of the Gaussian theory \eqref{eq:mOS_gauss}. For the RDLG, instead, $\Delta \simeq 1$ and $\beta/\nu \simeq 1/2$ and, once more, Eq. \eqref{eq:m_aspect} predicts $m \sim \sqrt{L_\perp/ L_\|},$ similar to the Gaussian theory,
in spite of the fact that the stationary state of the RDLG is definitely non-Gaussian. It is to be emphasized, however, that this behavior is expected to hold only when an appropriate finite-size scaling is performed, \ie, when different lattice sizes are compared at fixed $S_{\Delta}$, with the proper value for $\Delta,$ which is different for IDLG and RDLG (see Table \ref{tab:d2}). 

\begin{figure}[t]
 \centering
 \includegraphics[width=8.8 cm]{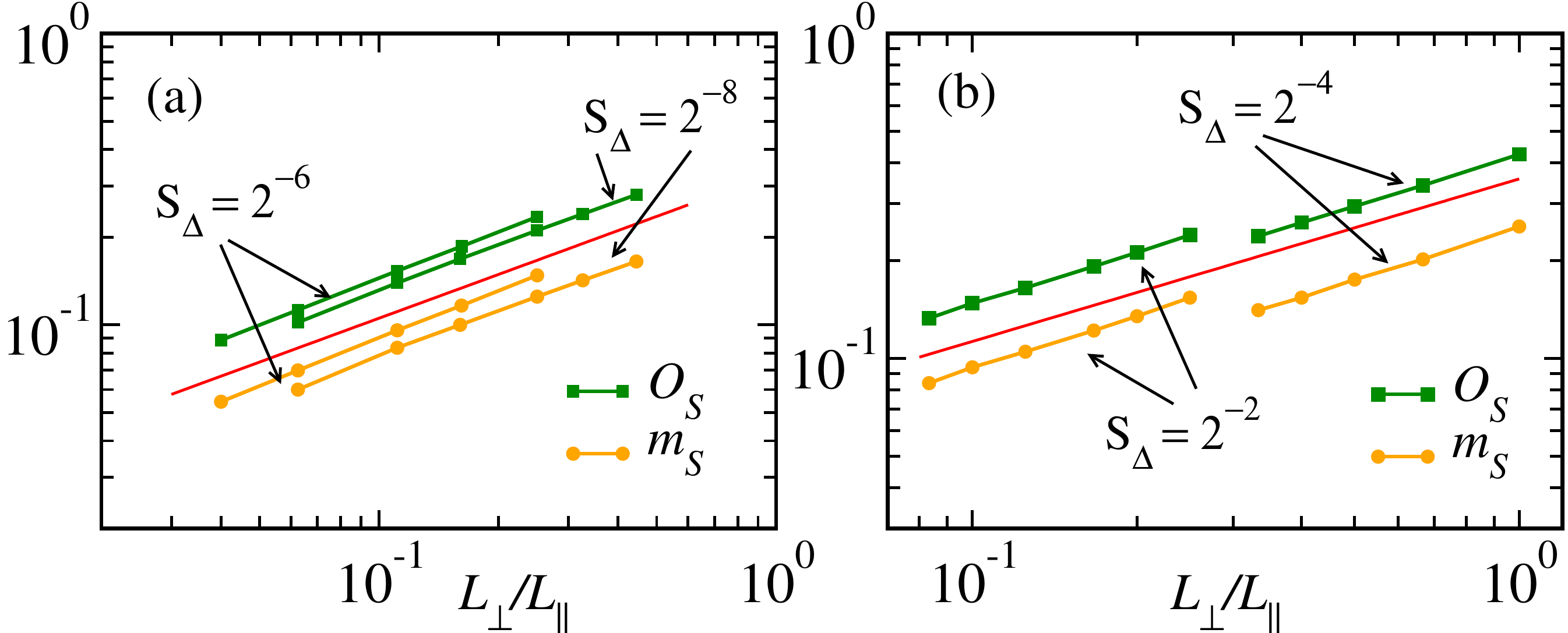} 
 \caption{Stationary values of $O$ and $m$ as functions of  $L_\perp/L_{||}$: (a) IDLG with $\Delta=2$ and two different values of $S_\Delta=2^{-8}$ and $2^{-6}.$  (b) RDLG with $\Delta =1$ and $S_\Delta=2^{-2}$ and $2^{-4}.$ The solid red lines indicate the dependence $\sqrt{L_\perp/L_{||}},$ as predicted by the Gaussian theory.}
 \label{fig:stat_idlg}
\end{figure}

Figure \ref{fig:stat_idlg} shows plots of the stationary values  $m_S$ and $O_S$  as functions of $L_\perp/L_{||}$ for IDLG with $\Delta =2$  [panel (a)] and RDLG with $\Delta =1$ [panel (b)], for two sets of values of $S_\Delta.$ The behavior of both the observables are consistent with the prediction in Eq.~\eqref{eq:m_aspect}. Accordingly, the dependence of the stationary values of the order parameters on  the isotropic aspect ratio, cannot be used in order to distinguish between the IDLG and the RDLG universality classes \footnote{Note, however, that if the same quantities are plotted as functions of only the orthogonal lattice size $L_\perp,$  the two models are expected to be distinguishable according to  Eq.~\eqref{eq:mL_scaling}.}.

Note that, it is not possible to make a similar analysis for the LG since $\Delta =0$ in this case and one cannot vary the aspect ratio keeping $S_\Delta$ fixed.

\subsection{Stationary values of the Binder cumulant}\label{sec:tcrdlg}

An effective and direct way of distinguishing between the different universality classes of lattice gas models is to investigate the Binder cumulants in the stationary state \cite{Caracciolo1, prl-dlg}. In the thermodynamic limit, the stationary value of Binder cumulant defined in Eq. \eqref{eq:binder} vanishes at the critical temperature in the case of the IDLG (consistently with a Gaussian behavior) while it converges to a value independent of the system size for the RDLG and the LG. Once again, these scaling behaviors are observed as long as the finite-size scaling is performed at fixed $S_\Delta$, with the proper value of $\Delta$ for the different models. 

\begin{figure}[t]
 \centering
 \includegraphics[width=6 cm]{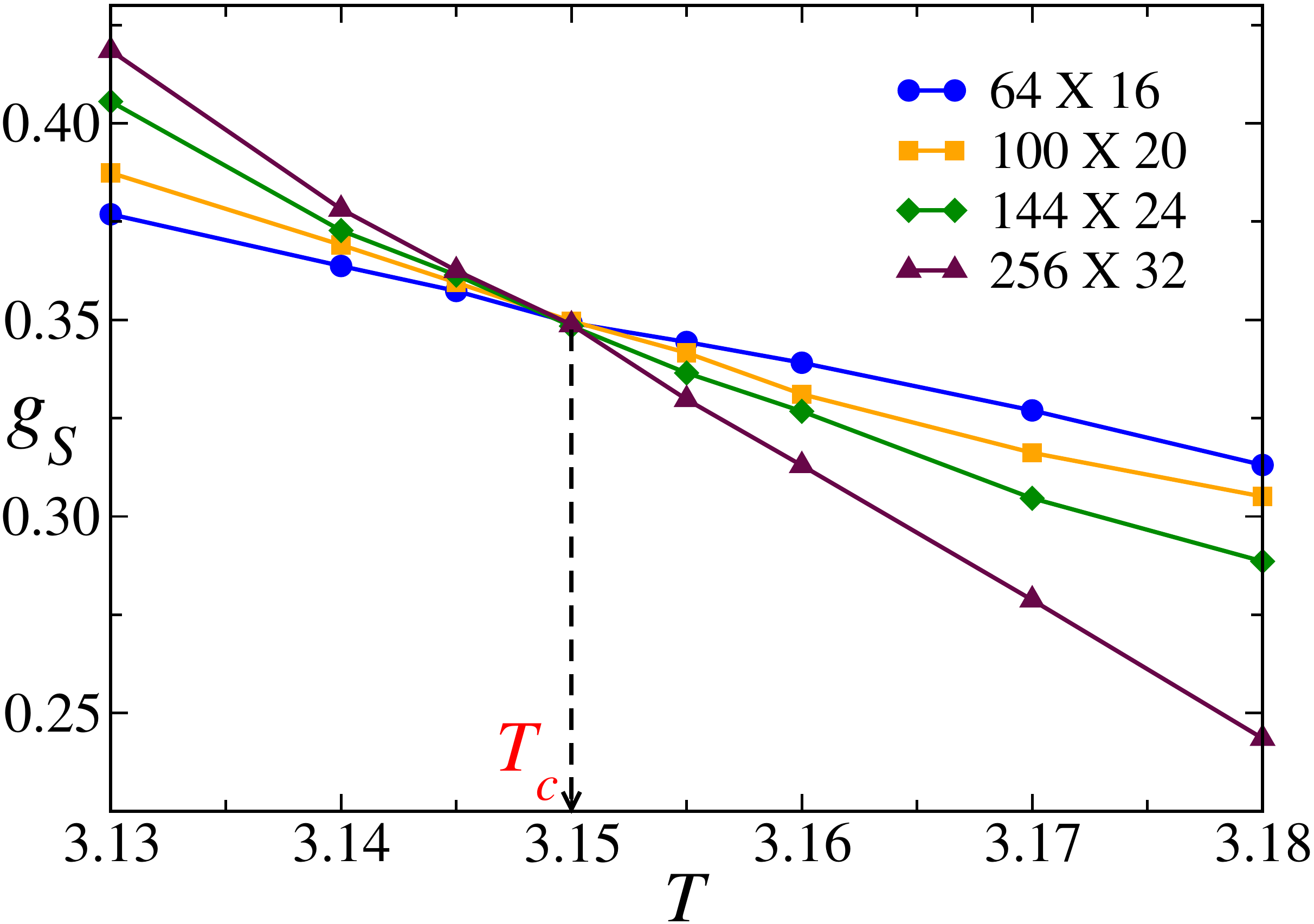}
 \caption{Determination of $T_c^\text{RDLG}$: Stationary value $g_S$ of the Binder cumulant is plotted as a function of the temperature $T$ for a set of lattice sizes $L_\| \times L_\perp$  with fixed $S_\Delta = 2^{-2}$ and $\Delta=1.$}
 \label{fig:stat_rdlg}
\end{figure}

The finite-size scaling behavior of the Binder cumulant is also widely used to determine the value of the critical temperature in various equilibrium systems \cite{Binder}. The method relies on the fact that, for certain systems, including LG,  the Binder cumulant attains a stationary value at the critical point, which does not depend on the system size. Here we use this fact in order to determine the critical temperature of the RDLG. Figure  \ref{fig:stat_rdlg} shows the plot of the stationary value $g_S$ of the Binder cumulant  as a function of the temperature $T$ for different geometries, but with a fixed $S_\Delta = 2^{-2}$ for $\Delta=1.$ The crossing point of the curves provides an accurate estimate of the critical temperature $T_c^{\text{RDLG}} = 3.150(5).$ This value has been used throughout this work and also in Ref.~\cite{prl-dlg}. 
 
\section{Conclusions}\label{sec:concl}

In this work we have investigated the critical behavior of conserved lattice gas models, both driven and undriven, and the possibility to describe it with an effective Gaussian theory. The three models studied here, namely, the infinitely strongly driven lattice gas (IDLG), the randomly driven lattice gas (RDLG) and the equilibrium lattice gas (LG), belong to three different universality classes. However, in Ref.~\cite{prl-dlg} it was shown that in the short-time dynamics after a critical quench, all these conserved lattice gases behave in a similar way, which is consistent with the transverse modes being described by a Gaussian theory. Hence the dynamics of transverse observables in this regime cannot be used to infer the universality classes of the different models. Here we elaborate and substantially extend the ideas and results anticipated in Ref.~\cite{prl-dlg}, providing additional analytical results and numerical evidence. 

The phase transitions in these conserved driven lattice gas models are characterized by considering the behavior of the order parameters which are ``transverse'' in nature, meaning that they are insensitive to fluctuations along the direction of the drive. The dynamics  of two of such order parameters $m$ and $O$ (see Eqs.~\eqref{eq:mdef} and \eqref{eq:OP}), and the auto-correlation of $m$ (see Eq.~\eqref{eq:def-auto}) are then predicted analytically assuming that the transverse modes are effectively described by a Gaussian theory. 

These predictions of the Gaussian theory compared with the results of Monte Carlo simulations from which it emerges that the dynamical behavior of both the order parameters $m$ and $O$  and the auto-correlation of $m$ agree very well with the Gaussian theory, in all the three different models, with the sole exception of $O$ in LG (see Fig.~\ref{fig:growth} and \ref{fig:Cm}). On this basis a unified short-time behavior emerges for the driven and undriven lattice gases, irrespective of the fact that their critical behaviors actually belong to different universality classes. 

The Gaussian theory also provides a way to determine normalization constants which appear to be arbitrary in the usual field-theoretic description. We extract the effective coarse-grained diffusion constant $\alpha$ and the effective temperature $T_\eta$ associated with the noise from the short-time behavior of the two order parameters in the driven lattice gases. The value of $\al$  and $T_\eta,$ for the IDLG and the RDLG, turn out to be almost equal. For the LG, however, one can only determine the product $\al T_\eta$, and the corresponding value turns out to be quite different from that for IDLG and RDLG.

To investigate the origin of the short-time Gaussian behavior we have studied the dynamical evolution  of the Binder cumulant $g$ starting from various initial conditions, including the fully disordered one. It appears that, as long as the order parameter vanishes in the initial configuration, the Binder cumulant signals a Gaussian-like behavior for a considerably long duration  as it remains very close to zero (see Fig.~\ref{fig:init}). 


We have also studied the behavior of $m,O$ and $g$ in these models in the stationary state. In contrast to the short-time regime, the stationary state bears the signatures of the specific critical behavior.  In fact, the stationary behavior of $g$ is very different in the three different lattice gases investigated here. We exploit the finite-size behavior of the stationary value of $g$ in the  RDLG in order  to determine accurately  the corresponding critical temperature.  
However, it turns out that, even in the stationary state, the dependence of $m$ and $O$ on the isotropic aspect ratio does not discriminate between the IDLG and RDLG universality classes.

In summary, we have shown that the short-time behavior of the transverse observables in the lattice gas models, both driven and undriven, is described by an effective Gaussian theory irrespective of them belonging to different universality classes. The origin of this ``super-universal'' behavior in the short-time regime may be related with the presence of a local density conservation in all these systems, which slows down the dynamics considerably \cite{prl-dlg}. However, the various models display their distinct critical behaviors in the stationary states.   Our work emphasizes the importance of the choice of order parameters, particularly in the presence of conservation laws.

\appendix

\section{Perturbative calculations}\label{sec:pert}

The critical behavior of the equilibrium LG is known to be described by an isotropic $\phi^4$ effective theory. However, the behavior of the Binder cumulant $g$ measured in Monte Carlo simulations (see Fig. \ref{fig:init}) is seemingly consistent with the Gaussian theory in the short-time regime following a critical quench. To understand this surprising fact, we perform a first-order perturbative calculation of the Binder cumulant for the  LG. For simplicity, we assume the lattice to be sufficiently large to replace it with a continuum. 

The time evolution of the coarse-grained spin-field $\phi(x,t)$ in LG is governed by the Langevin equation \eqref{eq:LG-phi} where $u$ denotes the strength of the perturbation. To linear order in $u,$
\beq
\phi(x,t)=\phi^0(x,t)+ u\,\phi^1(x,t)
\eeq
where $\phi^0(x,t)$ is the solution of Eq.~\eqref{eq:LG-phi} with $u=0$ (Gaussian) and $\phi^1(x,t)$  is the perturbative correction. It is useful to recall that the Fourier transform $\phi^0_{k}(t)$ of  $\phi^0(x,t)$ is the continuum version of Eq. \eqref{eq:gauss_soln}, \ie,
\bea
\phi^0_{k}(t)= ik \int_0^t \id s~ \eta_k(s) e^{-\gamma_k(t-s)} \label{eq:gauss_cont}
\eea
where $\gamma_k = \al k^2(\tau + k^2)$ (see Eq.~\eqref{eq:gammak} with $\hat k \to k$), and $\eta_k$ is the white noise on the continuum with $\la \eta_k(t) \eta_{k'}(t') \ra = 2 \al T_\eta (2 \pi)^d \delta(k+k')\delta(t-t').$
The time evolution of the Fourier transform $\phi^1_k(t)$ of the linear correction $\phi^1(x,t)$ follows from Eq. \eqref{eq:LG-phi} which takes the form
\bea
 \frac{d}{dt} \phi_k^1(t) = -\gamma_k \phi^1_k(t) - \alpha k^2 f_k(t), \label{eq:phi1k}
\eea
where $f_k(t)$ is the Fourier transform of $[\phi^0(x,t)]^3,$ $i.e.,$
\bea
 f_k(t)&=& \int \frac{\id k_1}{(2 \pi)^d} \frac{\id k_2}{(2 \pi)^d} ~ \phi^0_{k-k_1}(t) \phi^0_{k_1-k_2}(t) \phi^0_{k_2}(t).~~ \label{eq:fkt}
\eea
Eq.~\eqref{eq:phi1k} is can be solved and yields,
\bea
\quad \phi_k^1(t)&=& - \alpha k^2 \int_0^t \id s ~ e^{-\gamma_k(t-s)} f_k(s). \label{eq:phi1k_soln}
\eea
To compute the Binder cumulant of the $k$-th mode, as defined in Eq. \eqref{eq:binder},  we need to evaluate the  second and  fourth moment of $\phi_k(t).$ 
For the Gaussian theory, the Binder cumulant $g$ vanishes while, to the leading order in $u$ it takes the value
\bea
g = u \left [\frac {4 \delta_2}{ \la |\phi_k^0|^2 \ra } - \frac {\delta_4}{ \la |\phi_k^0|^2 \ra^2} \right] \label{eq:gu}
\eea
where the coefficients $\delta_2$ and $\delta_4$ are correlations between the Gaussian field $\phi^0$ with the linear correction $\phi^1$: 
\bea
\delta_2 &=& \la \phi_k^0 \phi_{-k}^1 \ra + \la \phi_{-k}^0 \phi_{k}^1 \ra \n \\ [0.25 em]
\delta_4 &=& 2 [ \la \phi_k^0 \phi_{-k}^0 \phi_{-k}^0 \phi_{k}^1 \ra + \la \phi_k^0 \phi_{k}^0 \phi_{-k}^0 \phi_{-k}^1 \ra].  \label{eq:del4}
\eea
Here all the fields $\phi_k^0$ and $\phi_k^1$ are evaluated at the same time $t.$ Since $\phi_k^1$ contains product of three $\phi_k^0$s, $\delta_2$ and $\delta_4$ are four- and six-point correlations of the Gaussian field which can be evaluated via Wick's theorem. The two contributions in $\delta_4$ are connected by a $k \to -k$ exchange and it is straightforward to see that   
\bea
\delta_4 = 4 \langle |\phi_k^0|^2 \rangle \delta_2 -4 \tilde g, \label{eq:del4_2}
\eea
where $\tilde g$ contains the contributions from the ``connected'' terms, \ie, terms in which each $\phi_k^0(t)$ is contracted with a $\phi_{k'}^0(s)$ for a time $t>s.$ There are six such connected contributions which can be  obtained explicitly from Eqs.~\eqref{eq:fkt} and \eqref{eq:phi1k_soln}. It is easy to see that they all contribute the same and we finally get
\bea
  \tilde g &=& 6\al k^2 \int_0^t \id s~ e^{-\gamma_k(t-s)} \int \frac{\id k_1}{(2\pi)^{d}} \frac{\id k_2} {(2\pi)^{d}}\,  \cr 
   && \times \langle \phi_k^0(t) \phi_{k-k_1}^0(s)\rangle  \langle \phi_{-k}^0(t) \phi^0_{k_1-k_2}(s)\rangle\langle \phi_{-k}^0(t) \phi_{k_2}^0(s)\rangle. \cr
&& \label{eq:gtilde1}
 \eea
The auto-correlation of the Gaussian field $\phi_k^0(t)$ is directly obtained from Eq.~\eqref{eq:gauss_cont}, 
\bea
\langle \phi_k^0(t) \phi_{k'}^0(s)\rangle &=&  \al T_\eta \frac{(2 \pi)^d k^2}{\gamma_k} e^{-\gamma_{k}(t-s)}(1- e^{-2\gamma_{k}s}) \delta (k+k') \cr
&& \label{eq:phi0ts}
\eea
for $t>s.$ Using Eq. \eqref{eq:phi0ts}, the momentum integrals in  Eq. \eqref{eq:gtilde1} can be calculated, and, at the critical point (\ie, with  $\gamma_k=\al k^4$) we get,
\bea 
  \tilde g 
&=& \frac{6 \alpha T_\eta^3 (2\pi)^{d}}{k^4} e^{-4 \alpha k^4 t} \int_0^t \id s ~e^{\alpha k^4 s} \left (e^{\alpha k^4 s} -e^{-\alpha k^4 s} \right)^3 \cr
&=& \frac{6 T_\eta^3 (2\pi)^{d}}{k^8}    \left[\frac{1}{4} + \frac{1}{2} e^{-6\al{k}^4t } + 3 e^{-4\al{k}^4 t} \left(\frac 14 + \al {k}^4 t \right)  \right. \cr
&& \qquad \qquad \quad \left.  - \frac{3}{2}e^{-2\al{k}^4t}  \right].  \label{eq:beta4_2}
   \eea

Finally, combining Eqs. \eqref{eq:gu}, \eqref{eq:del4_2}, and \eqref{eq:beta4_2} we get the linear correction to the Binder cumulant for a $\phi^4$ theory,
\bea
  g&=&\frac{4 u \tilde g}{\langle |\phi_k^0|^2\rangle ^2 } \cr 
&=&   \frac{24 uT_\eta}{(2\pi)^d k^4} \frac{1}{(1-e^{-2 \alpha k^4t})^2} \left[\frac{1}{4} + \frac{1}{2} e^{-6\al{k}^4t }  \right. \cr
&& \quad \quad \left. + 3 e^{-4\al{k}^4 t} \left(\frac 14 + \al {k}^4 t \right) 
  - \frac{3}{2}e^{-2\al{k}^4t} \right].\quad\label{eq:B_beta4}
\eea

We are particularly interested in the behavior of $g$ in the short-time regime,  which can be obtained by expanding the exponential in Eq. \eqref{eq:B_beta4} and by keeping the lowest order terms in $t.$ It turns out that the Binder cumulant grows quadratically upon increasing $t,$
\bea
g \sim k^4 t^2 + {\cal O}(t^3). \label{eq:g_cons}
\eea

In order to appreciate the role of the local  conservation of $\phi$ it is useful to repeat this calculation for the non-conserved field, in which case $\gamma_k = \al k^2$ at the critical point. Following the same steps, one finds,  
\bea
g \sim t^2 + {\cal O}(t^3).\label{eq:g_noncons}
\eea

This lowest order perturbative calculation is strictly valid around the upper critical dimensionality $d_c=4$ of the model. However, the qualitative feature that is of importance here is the fact that in the conserved case in Eq.~\eqref{eq:g_cons} the growth of $g$ is reduced by a factor of $k^4$ compared to the non-conserved case in Eq.~\eqref{eq:g_noncons}. For the first mode on a large lattice of linear size $L$, $k \sim 1/L$ and hence $g$ appears to be vanishingly small for the LG. This heuristic calculation provides a way to see how drastically conservation can alter the dynamical behavior of a system.


\begin{thebibliography}{1}%
\makeatletter
\providecommand \@ifxundefined [1]{%
 \@ifx{#1\undefined}
}%
\providecommand \@ifnum [1]{%
 \ifnum #1\expandafter \@firstoftwo
 \else \expandafter \@secondoftwo
 \fi
}%
\providecommand \@ifx [1]{%
 \ifx #1\expandafter \@firstoftwo
 \else \expandafter \@secondoftwo
 \fi
}%
\providecommand \natexlab [1]{#1}%
\providecommand \enquote  [1]{``#1''}%
\providecommand \bibnamefont  [1]{#1}%
\providecommand \bibfnamefont [1]{#1}%
\providecommand \citenamefont [1]{#1}%
\providecommand \href@noop [0]{\@secondoftwo}%
\providecommand \href [0]{\begingroup \@sanitize@url \@href}%
\providecommand \@href[1]{\@@startlink{#1}\@@href}%
\providecommand \@@href[1]{\endgroup#1\@@endlink}%
\providecommand \@sanitize@url [0]{\catcode `\\12\catcode `\$12\catcode
  `\&12\catcode `\#12\catcode `\^12\catcode `\_12\catcode `\%12\relax}%
\providecommand \@@startlink[1]{}%
\providecommand \@@endlink[0]{}%
\providecommand \url  [0]{\begingroup\@sanitize@url \@url }%
\providecommand \@url [1]{\endgroup\@href {#1}{\urlprefix }}%
\providecommand \urlprefix  [0]{URL }%
\providecommand \Eprint [0]{\href }%
\providecommand \doibase [0]{http://dx.doi.org/}%
\providecommand \selectlanguage [0]{\@gobble}%
\providecommand \bibinfo  [0]{\@secondoftwo}%
\providecommand \bibfield  [0]{\@secondoftwo}%
\providecommand \translation [1]{[#1]}%
\providecommand \BibitemOpen [0]{}%
\providecommand \bibitemStop [0]{}%
\providecommand \bibitemNoStop [0]{.\EOS\space}%
\providecommand \EOS [0]{\spacefactor3000\relax}%
\providecommand \BibitemShut  [1]{\csname bibitem#1\endcsname}%
\let\auto@bib@innerbib\@empty
\bibitem [{Note1()}]{Note1}%
  \BibitemOpen
  \bibinfo {note} {Note, however, that if the same quantities are plotted as
  functions of only the orthogonal lattice size $L_\perp ,$ the two models are
  expected to be distinguishable according to Eq.~\protect \textup {\hbox
  {\mathsurround \z@ \protect \normalfont (\ignorespaces \ref
  {eq:mL_scaling}\unskip \@@italiccorr )}}.}\BibitemShut {Stop}%
\end{thebibliography}%


\begin{thebibliography}{10}

\bibitem{Kadanoff} L. P. Kadanoff,
in {\it Critical Phenomena}, Proceedings of the Int. School of Physics, ``Enrico Fermi", Course LI, edited by M.S. Green, Academic Press, New York  (1971).

\bibitem{TauberBook} U. C. Täuber,  {\it Critical dynamics: a field theory approach to equilibrium and non-equilibrium scaling behavior}, Cambridge University Press (2014).

\bibitem{HalperinHohenberg1977} P. Hohenberg and B. I. Halperin, 
Rev. Mod. Phys. {\bf 49}, 435  (1977).

\bibitem{DP} J. Cardy  and U. C. T\"{a}uber, 
Phys. Rev. Lett.  {\bf 77}, 4780 (1996).

\bibitem{KPZ} M. Kardar,  G. Parisi, and Y. Zhang,
Phys. Rev. Lett.  {\bf 56}, 889 (1986).


\bibitem{JSS} H. K. Janssen, B. Schaub, and B. Schmittmann,
Z. Phys. B {\bf 73}, 539 (1989).

\bibitem{Zheng} B. Zheng,  
Int. J. Mod. Phys. B {\bf 12}, 1419 (1998).

\bibitem{ASreview}  E. V. Albano $et. al.,$  Rep. Prog. Phys. {\bf 74}, 026501 (2011).

\bibitem{AS2002} E. V. Albano and G. Saracco, Phys. Rev. Lett. {\bf 88}, 145701 (2002). 

\bibitem{Tauber} G. L. Daquila and U. C. T\"{a}uber, Phys. Rev. Lett. {\bf 108}, 110602 (2012).

\bibitem{Dickman} J. Marro and R. Dickman, {\it Nonequilibrium Phase Transitions in Lattice Models}, Cambridge University Press (2005).

\bibitem{DDSbook} B. Schmittmann and R. K. P. Zia, in {\it Statistical Mechanics
of Driven Diffusive Systems}, Phase Transitions and Critical Phenomena Vol. 17, edited by C. Domb and
J. L. Lebowitz, Academic Press, London (1995).

\bibitem{visco} H. Tanaka, J. Phys.: Condens. Matter {\bf 12}, R207 (2000).

\bibitem{traffic} D. Chowdhury, L. Santen, A. Schadschneider, Phys. Rep. {\bf 329}, 199 (2000).

\bibitem{active2} J. Stenhammar, A. Tiribocchi, R. J. Allen, D. Marenduzzo, and M. E. Cates,
Phys. Rev. Lett. {\bf 111}, 145702 (2013).

\bibitem{Cates} R. Wittkowski, A. Tiribocchi, J. Stenhammar, R. J. Allen, D. Marenduzzo, and M. E. Cates, Nature Communications {\bf 5}, 4351 (2014).


\bibitem{Mussardo} G. Mussardo, {\it Statistical field theory: an introduction to exactly solved models in statistical physics}, Oxford University Press (2010).

\bibitem{KLS} S. Katz, J. L. Lebowitz, and H. Spohn, Phys. Rev. B {\bf 28}, 1655(R) (1983); J. Stat. Phys. {\bf 34}, 497 (1984).

\bibitem{rdlg-S} B. Schmittmann, Europhys. Lett. {\bf 24}, 109 (1993).

\bibitem{prl-dlg}  U. Basu, V. Volpati, S. Caracciolo, and A. Gambassi, Phys. Rev. Lett. {\bf 118}, 050602 (2017).

 

\bibitem{Baxter} R. J. Baxter, {\it Exactly Solved Models in Statistical Mechanics}, Academic Press, London (1982).

\bibitem{JS} H. K. Janssen and B. Schmittmann, Z. Phys. B {\bf 64}, 503 (1986).

\bibitem{LC} K.-t. Leung and J. L. Cardy, J. Stat. Phys. {\bf 44}, 567 (1986).

\bibitem{Caracciolo1} S. Caracciolo, A. Gambassi, M. Gubinelli, and A. Pelissetto, J. Phys. A {\bf 36}, L315 (2003).

\bibitem{rdlg2} S. Caracciolo, A. Gambassi, M. Gubinelli, and A. Pelissetto, Phys. Rev. E {\bf 72}, 056111 (2005).

\bibitem{Kubo} R. Kubo, Rep. Prog. Phys. {\bf 29}, 255 (1966).

\bibitem{math-book} M. Abramowitz  and I. A. Stegun (Eds.), {\it Handbook of Mathematical Functions}, Dover, New York (1972).

\bibitem{multi-G} A. Gut, {\it An Intermediate Course in Probability}, Springer, New York (2009).

\bibitem{Binder} K. Binder and D. Heermann, {\it Monte Carlo Simulation in Statistical Physics}, Springer-Verlag, Berlin Heidelberg (2010).

\bibitem{Caracciolo2} S. Caracciolo, A. Gambassi, M. Gubinelli, and A. Pelissetto, J. Stat. Phys. {\bf 115}, 281 (2004).

\bibitem{Godreche} C. Godr\`{e}che,  F. Krzaka\l{}a, and F. Ricci-Tersenghi, J. Stat. Mech.: Theor. Exp.  P04007 (2004).



 



%




\end{thebibliography}
\end{document}